\documentclass[aps,floatfix,amsmath,nofootinbib,amssymb,superscriptaddress]{revtex4}

\usepackage{overpic}
\usepackage{amssymb}
\usepackage{indentfirst}
\usepackage{feynmf}   %{feynmp}
\usepackage{slashed}  %for Feynman symbols
\usepackage{cases}
\usepackage{color}
\usepackage{multirow}
\usepackage{epstopdf}
\usepackage{graphicx,color,bm}
\usepackage{epstopdf}

\usepackage[colorlinks,
            citecolor=green,
            anchorcolor=red,
            menucolor=red,
            linkcolor=red,
            filecolor=red,
            runcolor=red,
            urlcolor=blue,
            frenchlinks=red]{hyperref}

\begin{document}

\title{The quark quasi Sivers function and quasi Boer-Mulders function in a spectator diquark model}

\author{Chentao Tan}
\affiliation{School of Physics, Southeast University, Nanjing 211189, China}

\author{Zhun Lu}
\email{zhunlu@seu.edu.cn}
\affiliation{School of Physics, Southeast University, Nanjing 211189, China}

\begin{abstract}
We compute the leading-twist T-odd quasi-distributions of the proton in a spectator model with scalar and axial-vector diquarks: the quasi Sivers function $\tilde{f}_{1T}^\perp(x, \bm k_T^2; P_z)$ and the quasi Boer-Mulders function $\tilde{h}_1^\perp(x, \bm k_T^2; P_z)$. We obtain the quark-quark correlators in the four-dimensional Euclidian space by replacing $\gamma^+$ and $\sigma^{i+}$ in the light-cone frame with $\gamma_z$ and $\sigma_{iz}$. We show by analytical calculation that the results of $\tilde{f}_{1T}^\perp$ and $\tilde{h}_1^\perp$ derived from the correlators can reduce to the expressions of the corresponding standard T-odd distributions $f_{1T}^\perp(x, \bm k_T^2)$ and $h_1^\perp(x, \bm k_T^2)$ in the limit $P_z\rightarrow\infty$. The numerical results for these quasi-distributions and their first transverse moments for the $u$ and $d$ quarks in different $x$ and $P_z$ regions are also presented. We find that $\tilde{f}_{1T}^{\perp(1)}(x, P_z)$ and $\tilde{h}_1^{\perp(1)}(x, P_z)$  in the spectator model are fair approximations to the standard ones (within 20-30\%) in the region $0.1<x<0.5$ when $P_z \geq 2.5-3$ GeV. This supports the idea of using T-odd quasi-distributions to obtain standard distributions in the region $P_z>2.5$ GeV as fair approximation.
\end{abstract}

%\pacs{12.38.-t, 13.85.Qk, 13.88.+e}
\maketitle

\section{Introduction}

The Parton distribution functions (PDFs), defined through the light-cone correlation functions are of fundamental importance in hadronic physics.
They describe the density of a parton carrying in hadron a light-cone fraction $x$ of the total momentum.
Although PDFs are difficult to calculate from the first principle of QCD, they play crucial role in the description of various high energy inclusive processes via QCD factorization theorem.
A natural extension of PDFs is the transverse-momentum dependent distributions(TMDs)~\cite{Bacchetta:2006tn}.
They encode the probability density of a parton inside the nucleon with longitudinal momentum fraction $x$ and transverse momentum $\bm{k}_T$.
In leading-twist there are eight TMDs, corresponding to different polarization states of the hadron and the parton.
Of particular interests are two T-odd TMDs, namely, the Sivers function $f_{1T}^\perp(x, \bm k_T^2)$ and the Boer-Mulders function $h_1^\perp(x, \bm k_T^2)$.
The former one describes the asymmetric distribution of the unpolarized parton in a transverse polarized hadron~\cite{Sivers:1989cc,Anselmino:2005an}, while the latter one describes the distribution of the transverse polarized parton in an unpolarized hadron\cite{Boer:1997nt}.
For these reasons they can give rise to the spin or azimuthal asymmetries in semi-inclusive deep inelastic scattering process or Drell-Yan process.

Recently, the concept of quasi-PDFs for hadrons has been proposed in Refs.~\cite{Ji:2013dva, Ji:2014gla} and has received a lot of attention.
Different from the standard PDFs, quasi-PDFs are defined by the bilocal operators on a spatial interval such that they can be calculated by the lattice QCD in a four-dimensional Euclidian space.
Quasi-PDFs have a parton probability interpretation similar to the standard PDFs, but for a parton carrying a fraction $x$ of the finite momentum $\vec P$ of the hadron.
Although introducing quasi-PDFs will bring a explicit dependence on the hadron momentum (usually denoted by $P_z = |\vec P|$),
it is found that in large limit of $P_z$, the quasi-PDF $\tilde{f}(x,P_z)$ converges to the standard PDF $\tilde{f}(x)$.
This makes it possible to calculate the $x$-dependence of the standard PDFs by using Lattice QCD.
In particular, a number of lattice calculations on quasi-PDFs and related quantities~\cite{Lin:2014zya, Alexandrou:2015rja, Alexandrou:2016jqi, Chen:2016utp, Zhang:2017bzy, Zhang:2017zfe,Alexandrou:2017huk, Chen:2017mzz, Green:2017xeu, Lin:2017ani, Orginos:2017kos, Bali:2017gfr, Alexandrou:2017dzj, Chen:2017gck, Alexandrou:2018pbm, Chen:2018xof, Alexandrou:2018eet, Liu:2018uuj, Bali:2018spj, Lin:2018qky,LatticeParton:2018gjr} have been performed.

The framework of the quasi-distributions can be also extended to the case of TMDs, as already proposed in Ref.~\cite{Ji:2013dva}.
The quasi-TMD $\tilde{f}(x,k_T^2;P_z)$ has the parton probability interpretation similar to the TMD, but is defined in the Euclidean space and depends on the hadron momentum $P_z$.
In Refs.~\cite{Ji:2018hvs,Ebert:2019okf}, the basic procedure that can be used to compute the TMDs from lattice QCD using large momentum effective theory (LAMET)~\cite{Ji:2020ect} or quasi-TMDs has been laid out.
The T-even spin-dependent quasi-TMDs that are amenable to lattice QCD calculations and that can be used to determine standard spin-dependent TMDPDFs have also been constructed in Ref.~\cite{Ebert:2020gxr}.
Furthermore, the quark Sivers function is computed~\cite{Ji:2020jeb} in the leading-order expansion in the framework of LAMET.

%We have also known that these two TMDs are resulted from the initial-state interactions of the scattering %hadrons~\cite{Boer:2002ju}.
In this work, we will study the quasi-distributions of the T-odd TMDs from the model aspects.
As demonstrated in Refs.~\cite{Gamberg:2014zwa, Bacchetta:2016zjm, Nam:2017gzm, Broniowski:2017wbr, Hobbs:2017xtq, Broniowski:2017gfp, Xu:2018eii,Son:2019ghf}, model calculations on quasi-distributions can provide useful information for which values of $P_z$ the quasi-PDFs are fair approximations of standard PDFs.
To explore for what values of $P_z$ the T-odd quasi-TMDs and the standard T-odd TMDs are approximations of each other, we calculate the quark quasi Sivers function $\tilde{f}_{1T}^\perp(x,k_T^2;P_z)$ and quasi Boer-Mulders function $\tilde{h}_1^\perp$ using a spectator diquark model.
We would like to study the flavor-dependence of these quasi-distributions, therefore, in the calculation we include both the scalar diquark and the axial-vector diquark to obtain the distributions of $u$ and $d$ quarks.
In addition, we select the dipolar form factor for the proton-quark-diquark vertex.

This paper is organized as follows: In Sec.~\ref{Sec:2}, we present the definitions of standard T-odd TMDs and the corresponding quasi-TMDs by using the light-cone correlators and the Euclidean correlators, respectively.
In Sec.~\ref{Sec:3}, we perform the calculations of two quasi-distributions in the spectator model with scalar and axial-vector diquarks. In Sec.~\ref{Sec:4}, we give the numerical results for the quasi-functions and the first $\bm{k}_T$-moment of two functions to explore the dependence of these distributions on $x$, $P_z$, $\bm{k}_T$. We provide some conclusions in Sec.~\ref{Sec:5}.

\section{Definition: standard Sivers function and Boer-Mulders function, quasi Sivers function and quasi Boer-Mulders function} \label{Sec:2}

In this section, we present the operator definitions for the standard TMD distributions $f_{1T}^\perp$ and $h_1^\perp$, as well as the quasi-TMD-distributions $\tilde{f}_{1T}^\perp$ and $\tilde{h}_1^\perp$, respectively.

The standard TMD distributions are usually expressed in the light-cone coordinate, in which one writes
$a^{\pm}=(a^0\pm a^3)/\sqrt{2}$ and $\bm a_T=(a_1, a_2)$ for an arbitrary four-vector $a^\mu$ in a specific reference frame, and the components of $a^\mu$ are given as $(a^+, a^-, \bm a_T)$.
The standard TMD distributions for a quark with light-cone momentum fraction $x=k^+/P^+$ and transverse momentum $\bm k_T$ appear in the decomposition of the quark-quark correlation function $\Phi$ (in DIS)
\begin{align}
\Phi_{ij}(x,\bm{k}_T,S)=\int \frac{d\xi^-d^2\bm\xi_T}{(2\pi)^3} e^{-i \xi\cdot k} \langle P,S|\bar{\psi}_j(0) \mathcal{U}^{n^-}[0,+\infty] \mathcal{U}^{n^-}[+\infty,\xi] \psi_i(\xi) |P,S\rangle{\big{|}}_{\xi^+=0},
\end{align}
which can be parameterized according to the hermiticity, parity invariance and charge conjugation invariance.
In the above equation, $k^\mu$ is the momentum of the quark,
\begin{align}
P^\mu=(P^+, P^-, \bm{0}_T),~~~S^\mu=\left(S_L {P^+\over M}, -S_L {P^-\over M}, \bm{S}_T\right)
\end{align}
are the momentum and the polarization vector of the nucleon, respectively.
Furthermore, $\mathcal{U}^{n^-}$ are the gauge links to ensure the gauge invariance of the operator definition:
\begin{align}
\mathcal{U}^{n^-}[0,+\infty]&=\mathcal{P} \exp\left[-ig\int^{\infty^-}_{0^-} d\eta^- A^+(0^+,\eta^-,\bm{0}_T)\right] \mathcal{P}\exp\left[-ig\int^{\infty_T}_{0_T} d\zeta_T\cdot A_T(0^+,\infty^-,\zeta_T)\right],\\
\mathcal{U}^{n^-}[+\infty,\xi]&=
\mathcal{P}\exp\left[-ig\int^{\infty_T}_{\xi_T} d\zeta_T \cdot A_T(0^+,\infty^-,\zeta_T)\right]
\mathcal{P} \exp\left[-ig\int_{\xi^-}^{\infty^-} d\eta^- A^+(0^+,\eta^-,\xi_T)\right],
\end{align}
where $n^-=(0,1,\bm{0}_T)$, $\mathcal{P}$ denotes all possible ordered paths followed by the gluon field $A$, which couples to the quark field $\psi$ through the coupling constant $g$.

Then the standard Sivers function $f_{1T}^\perp(x,\bm{k}_T^2)$ and Boer-Mulders function $h_1^\perp(x,\bm{k}_T^2)$ can be defined by the following expressions~\cite{Bacchetta:2006tn}
\begin{align}
\frac{\epsilon_T^{ij}\bm{k}_{Ti}\bm{S}_{Tj}}{M}f_{1T}^\perp(x,\bm{k}_T^2)
&=-\frac{1}{4}\textrm{Tr}[\Phi^{[\gamma^+]}(x,\bm{k}_T,S)-\Phi^{[\gamma^+]}(x,\bm{k}_T,-S)]+\textrm{H.c.}
\label{eq:3},\\
\frac{\epsilon_T^{ij}\bm{k}_{Ti}}{M}h_1^\perp(x,\bm{k}_T^2)
&=\frac{1}{4}\textrm{Tr}[\Phi^{[i\sigma^{i+}\gamma_5]}(x,\bm{k}_T,S)
+\Phi^{[i\sigma^{i+}\gamma_5]}(x,\bm{k}_T,-S)]+\textrm{H.c.}
\label{eq:4},
\end{align}
where H.c. denotes the Hermitian conjugate terms, and
\begin{align}
\Phi^{[\gamma^+]}(x,\bm{k}_T,S)&={1\over 2} \textrm{Tr}\left[\Phi(x,\bm{k}_T,S) \gamma^+\right]
\label{eq:1},\\
\Phi^{[i\sigma^{i+}\gamma_5]}(x,\bm{k}_T,S)&={1\over 2} \textrm{Tr}\left[\Phi(x,\bm{k}_T,S) i\sigma^{i+}\gamma_5 \right]
\label{eq:2}.
\end{align}

On the other hand,  quasi-TMDs are defined as the matrix elements of the following equal-time spatial correlation function~\cite{Ji:2013dva}
\begin{align}
\tilde{\Phi}_{ij}(x,\bm{k}_T,S;P_z)=\int \frac{d\xi^z d^2\bm\xi_T}{(2\pi)^3} e^{-i \xi\cdot k} \langle P,S|\bar{\psi}_j(0) \mathcal{U}^{n_z}[0,+\infty] \mathcal{U}^{n_z}[+\infty,\xi] \psi_i(\xi) |P,S\rangle{\big{|}}_{\xi_0=0}, \label{eq:quasicorr}
\end{align}
where $x=k_z/P_z$ is the longitudinal momentum fraction of the quark, $\mathcal{U}^{n_z}[0,+\infty]$ and  $\mathcal{U}^{n_z}[+\infty,\xi]$ are the gauge links having the forms~\footnote{For a generic four-vector $a^\mu$, we denote the ordinary Minkowski components by $(a^0, \bm{a}_T, a^z)$}
\begin{align}
\mathcal{U}^{n^z}[0,+\infty]&=\mathcal{P} \exp\left[-ig\int^{\infty^z}_{0^z} d\eta^z A^z(0^0,\eta^z,\bm{0}_T)\right] \mathcal{P}\exp\left[-ig\int^{\infty_T}_{0_T} d\zeta_T\cdot A_T(0^0,\infty^z,\zeta_T)\right],\\
\mathcal{U}^{n^z}[+\infty,\xi]&=
\mathcal{P}\exp\left[-ig\int^{\infty_T}_{\xi_T} d\zeta_T \cdot A_T(0^0,\infty^z,\zeta_T)\right]
\mathcal{P} \exp\left[-ig\int_{\xi^z}^{\infty^z} d\eta^z A^z(0^0,\eta^z,\xi_T)\right].
\end{align}

Using the correlation function in Eq.~(\ref{eq:quasicorr}), we can write the expressions for calculating the quasi Sivers function $\tilde{f}_{1T}^\perp(x,\bm{k}_T^2,P_z)$ and the quasi Boer-Mulders function $\tilde{h}_1^\perp(x,\bm{k}_T^2,P_z)$ as follows ~\cite{Ji:2013dva,Gamberg:2014zwa}
\begin{align}
\frac{\epsilon_T^{ij}\bm{k}_{Ti}\bm{S}_{Tj}}{M}\tilde{f}_{1T}^\perp(x,\bm{k}_T^2,P_z)&=
-\frac{1}{4}\textrm{Tr}[\tilde{\Phi}^{[\gamma_z]}(x,\bm{k}_T,S)-\Phi^{[\gamma_z]}(x,\bm{k}_T,-S)]+H.c.
\label{eq.quasi-siv},\\
\frac{\epsilon_T^{ij}\bm{k}_{Ti}}{M}\tilde{h}_1^\perp(x,\bm{k}_T^2,P_z)
&=\frac{1}{4}\textrm{Tr}[\tilde{\Phi}^{[i\sigma_{iz}\gamma_5]}(x,\bm{k}_T,S)+\Phi^{[i\sigma_{iz}\gamma_5]}(x,\bm{k}_T,-S)]+H.c.
\label{eq.quasi-bm},
\end{align}
where $\tilde{\Phi}^{[\gamma_z]}$ and $\tilde{\Phi}^{[i\sigma_{iz}\gamma_5]}$ are defined as
\begin{align}
\tilde{\Phi}^{[\gamma_z]}(x,\bm{k}_T,S)&={1\over 2} \textrm{Tr}\left[\tilde{\Phi}(x,\bm{k}_T,S;P_z) \gamma_z\right]
\label{eq:quasi_gz},\\
\tilde{\Phi}^{[i\sigma_{iz}\gamma_5]}(x,\bm{k}_T,S)&={1\over 2} \textrm{Tr}\left[\tilde{\Phi}(x,\bm{k}_T,S;P_z) i\sigma^{iz}\gamma_5 \right]
\label{eq:guasi_g5}.
\end{align}

\section{Analytic calculation} \label{Sec:3}

In this section, we present the analytic calculation of the quasi Sivers function and the quasi Boer-Mulders function using a spectator model.

\begin{figure}
  \centering
  \includegraphics[width=0.6\columnwidth]{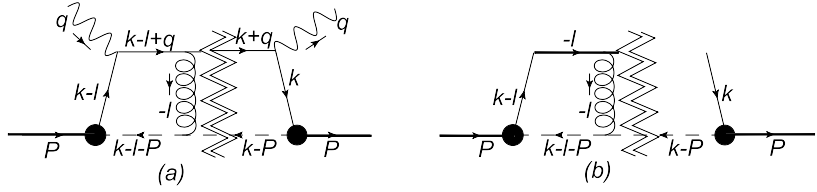}
  \caption{(a): Interference between the one-gluon exchange diagram and the tree-level diagram in the spectator diquark model. (b): Interference between the one-gluon exchange diagram and the tree-level diagram in the spectator diquark model in eikonal approximation.}
  \label{fig:int}
\end{figure}

The model has been widely used to calculate the standard TMDs~\cite{Lu:2004au,Kang:2010hg,Bacchetta:2003rz} and GPDs of the nucleon and the spin-0 hadron.
Recently it was also applied to calculate  quasi-PDFs~\cite{Gamberg:2014zwa} and quasi-GPDs~\cite{Bhattacharya:2018zxi,Bhattacharya:2019cme,Ma:2019agv}.
In the model, the nucleon is viewed as a two-body composite system of an active quark with mass $m$ and a diquark with mass $M_X$. The latter one can be a scalar diquark or an axial-vector diquark according to its spin. As shown in Fig.~(\ref{fig:int}), The proton-quark-diquark coupling is characterized by some effective vertices.
For this purpose we adopt the vertices for the scalar and the axial-vector diquarks as
\begin{align}
\textrm{scalar\ diquark}:\ &ig_s(k^2),\\
\textrm{axial-vector\ diquark}:\ &i\frac{g_a(k^2)}{\sqrt{2}}\gamma^\mu \gamma_5,
\end{align}
where $g_X(k^2)$ denotes the form factors of the coupling. In our calculations, we will use the dipolar form factor
\begin{align}
g_X(k^2)=g_X\frac{k^2-m^2}{|k^2-\Lambda_X^2|^2}=
g_X\frac{(k^2-m^2)(1-x)^2}{(\bm{k}_T^2+L_X^2(\Lambda_X^2))^2},
\end{align}
where
\begin{align}
L_X^2(\Lambda_X^2)=xM_X^2+(1-x)\Lambda_X^2-x(1-x)M^2,
\end{align}
with $(P-k)^2=M_X^2$. $g_X$ and $\Lambda_X$ denote the coupling constants and the cutoffs, respectively. They are considered as the free parameters of the model together with the diquark mass $M_X$.
In addition, the propagators of the scalar and axial-vector diquarks are given by
\begin{align}
scalar\ diquark:\ &\frac{i}{k^2-M_s^2},\\
axial-vector\ diquark:\ &\frac{i}{k^2-M_a^2}d^{\mu\nu}.
\end{align}
In this work, we adopt $d^{\mu\nu}=-g^{\mu\nu}$ to simplify the calculation. We admit that this polarization sum contains unphysical polarization states of the axial-vector diquark.

\subsection{Standard Sivers function and Boer-Mulders function}

The quark Sivers function and quark Boer-Mulders function have been calculated by various models in the literatures, such as the spectator model~\cite{Brodsky:2002cx, Boer:2002ju, Gamberg:2003ey, Bacchetta:2003rz,Lu:2004hu,Gamberg:2007wm,Bacchetta:2008af}, the light-cone quark model~\cite{Lu:2004au, Pasquini:2010af}, the non-relativistic constituent quark model~\cite{Courtoy:2008vi,Courtoy:2009pc}, and the MIT bag model~\cite{Yuan:2003wk,Courtoy:2008dn,Courtoy:2009pc}.
In order to make a comparison, we briefly review the procedure of calculating the T-odd TMDs in the spectator model.
To do this we expand the gauge-link to one loop order.
So the Sivers and Boer-Mulders functions can be computed from
\begin{align}
\frac{\epsilon_T^{ij}\bm{k}_{Ti}\bm{S}_{Tj}}{M}f_{1T}^\perp(x,\bm{k}_T^2)
&=-\frac{1}{4}\frac{1}{(2\pi)^3}\frac{1}{2(1-x)P^+}\textrm{Tr}\left[\left(\mathcal{M}^{(1)}(S)\bar{\mathcal{M}}^{(0)}(S)
-\mathcal{M}^{(1)}(-S)\bar{\mathcal{M}}^{(0)}(-S)\right)\gamma^+\right]+\textrm{H.c.},
\label{eq,9}\\
\frac{\epsilon_T^{ij}\bm{k}_{Ti}}{M}h_1^\perp(x,\bm{k}_T^2)
&=\frac{1}{4}\frac{1}{(2\pi)^3}\frac{1}{2(1-x)P^+}\textrm{Tr}\left[\left(\mathcal{M}^{(1)}(S)\bar{\mathcal{M}}^{(0)}(S)
+\mathcal{M}^{(1)}(-S)\bar{\mathcal{M}}^{(0)}(-S)\right)i\sigma^{i+}\gamma_5\right]+\textrm{H.c.},
\end{align}
where $\mathcal{M}^{(0)}$ and $\mathcal{M}^{(1)}$ are the tree level and one-loop level amplitudes of $p\rightarrow q X$ shown in Fig.~\ref{fig:int}(a). Now, we perform the so-called ``eikonal approximation" and take into account only the leading parts of the momenta of the quark after the photon scattering. Therefore, the eikonal propagator of the quark in the light-cone framework in Fig.~\ref{fig:int}(b) is given by
\begin{align}
\frac{i(\slashed{k}+\slashed{q}-\slashed{l}+m)}{(k+q-l)^2-m^2+i\epsilon}\approx \frac{i(k+q)^-\gamma^+}{-2l^+(k+q)^-+i\epsilon}=\frac{i}{2}\frac{\gamma^+}{-l^++i\epsilon}.
\end{align}
Then we have the expressions
\begin{align}
\mathcal{M}^{(0)}(S)&=\frac{i(\slashed{k}-m)}{k^2-m^2} ig_s(k^2) \frac{1+\gamma_5\slashed{S}}{2}U(P,S)
\label{eq,5},\\
\mathcal{M}^{(1)}(S)&=\int \frac{d^4l}{(2\pi)^4} \frac{i^2 e_c^2 \Gamma_{s\rho} n_-^\rho (\slashed{k}-\slashed{l}+m) ig_s((k-l)^2) }{(l^2-m_g^2+i\epsilon)(-l^++i\epsilon)((k-l)^2-m^2+i\epsilon)((P-k+l)^2-M_s^2+i\epsilon)} \frac{1+\gamma_5\slashed{S}}{2} U(P,S)
\label{eq,6},
\end{align}
for scalar diquark, and
\begin{align}
\mathcal{M}^{(0)}(S)=& \frac{i(\slashed{k}-m)}{k^2-m^2} \epsilon_\mu^*(P-k,\lambda_a) i\frac{g_a(k^2)}{\sqrt{2}} \gamma^\mu \gamma_5 \frac{1+\gamma_5\slashed{S}}{2} U(P,S)
\label{eq,7},\\
\mathcal{M}^{(1)}(S)=&\int \frac{d^4l}{(2\pi)^4} \frac{-i^2 e_c^2 \epsilon_\sigma^*(P-k,\lambda_a) \Gamma_{a\rho}^{\nu\sigma} n_-^\rho(\slashed{k}-\slashed{l}+m)d_{\mu\nu} }{(l^2-m_g^2+i\epsilon)(-l^++i\epsilon)((k-l)^2+i\epsilon)((P-k+l)^2-M_a^2+i\epsilon)}i\frac{g_a((k-l)^2)}{\sqrt{2}}\gamma^\mu \gamma_5\\
 &\times \frac{1+\gamma_5\slashed{S}}{2} U(P,S)
\label{eq,8},
\end{align}
for axial-vector diquark, where
\begin{align}
\Gamma_{s\rho} &=(2P-2k+l)_\rho,\\
\Gamma_{a\rho}^{\nu\sigma} &=(2P-2k+l)_\rho g^{\nu\sigma}-(P-k+(1+\kappa_a)l)^\sigma g^\nu_\rho-(P-k-\kappa_al)^\nu g^\sigma_\rho,
\end{align}
and $e_c$ denotes the color charge of the quark or diquark. $\kappa_a$ denotes the diquark anomalous chromomagnetic moment. 
Here we adopt $\kappa_a=0$.

Integrating over the loop momentum $l^\mu$, we arrive at the analytic expressions for the Sivers function and the Boer-Mulders function contributed by the scalar/axial-vector diquark components:
\begin{align}
f_{1T}^{\perp (s)}(x,\bm{k}_T^2)&=h_1^{\perp q(s)}(x,\bm{k}_T^2)=-\frac{g_s^2 e_c^2}{4(2\pi)^4}\frac{(1-x)^3(m+xM)M}{L_s^2(\Lambda_s^2)[\bm{k}_T^2+L_s^2(\Lambda_s^2)]^3},\\
f_{1T}^{\perp (a)}(x,\bm{k}_T^2)&=\frac{g_a^2 e_c^2}{8(2\pi)^4}\frac{(1-x)^2x(m+M)M}{L_a^2(\Lambda_a^2)[\bm{k}_T^2+L_a^2(\Lambda_a^2)]^3},\\
h_1^{\perp (a)}(x,\bm{k}_T^2)&=\frac{g_a^2 e_c^2}{8(2\pi)^4}\frac{(1-x)^2[(m+(2x-3)M)x-2m]M}{L_a^2(\Lambda_a^2)[\bm{k}_T^2+L_a^2(\Lambda_a^2)]^3}.
\end{align}

We can also compute the first transverse-moment of the two functions
\begin{align}
f_{1T}^{\perp (s)(1)}(x)&=h_1^{\perp q(s)(1)}(x)=\int d^2\bm{k}_T \frac{\bm{k}_T^2}{2M^2}f_{1T}^{\perp (s)}(x,\bm{k}_T^2)=-\frac{g_s^2 e_c^2}{32(2\pi)^3M}\frac{(m+xM)(1-x)^3}{[L_s^2(\Lambda_s^2)]^2},\\
f_{1T}^{\perp (a)(1)}(x)&=\int d^2\bm{k}_T \frac{\bm{k}_T^2}{2M^2}f_{1T}^{\perp (a)}(x,\bm{k}_T^2)=\frac{g_a^2 e_c^2}{64(2\pi)^3M}\frac{x(m+M)(1-x)^2}{[L_a^2(\Lambda_a^2)]^2},\\
h_1^{\perp (a)(1)}(x)&=\int d^2\bm{k}_T \frac{\bm{k}_T^2}{2M^2}h_{1}^{\perp (a)}(x,\bm{k}_T^2)=\frac{g_a^2 e_c^2}{64(2\pi)^3M}\frac{(1-x)^2[(m+(2x-3)M)x-2m]}{L_a^2(\Lambda_a^2)^2}.
\end{align}
Finally, we apply the following spin-flavor relation to obtain the distributions for the $u$ and $d$ valence quarks~\cite{Jakob:1997wg,Bacchetta:2003rz}:
\begin{align}
{f}^{u} = {3\over 2} {f}^{(s)} + {1\over 2} {f}^{ (a)}, ~~~~~ {f}^{ d} = {f}^{ (a)} , \label{eq:ud}
\end{align}
where $f$ can be $f_{1T}^\perp$ or $h_{1}^\perp$.

\subsection{The quasi Sivers and quasi Boer-Mulders functions}

Using Eqs.~(\ref{eq.quasi-siv}) and (\ref{eq.quasi-bm}) and Fig.~\ref{fig:int}, we can calculate the T-odd quasi-TMDs in a similar way. The main difference is the Feynman rules for the eikonal propagator and the eikonal vertex, for which they have the replacements:
 \begin{align}
 {i\over -l^+ +i\epsilon}& \ \rightarrow \ {i\over -l_z+i\epsilon}, \nonumber\\
-ie_c n_-^\mu & \ \rightarrow \-ie_c n_z^\mu\nonumber.
 \end{align}
Then we express $\tilde{\Phi}(x,\bm{k}_T,S;P_z)$ as
\begin{align}
\tilde{\Phi}(x,\bm{k}_T,S,P_z)=\frac{1}{(2\pi)^3}\frac{1}{2\lambda}(\mathcal{M}^{(1)}(S)\bar{\mathcal{M}}^{(0)}(S)\pm\mathcal{M}^{(1)}(-S)\bar{\mathcal{M}}^{(0)}(-S))
\label{eq.14}.
\end{align}
Here, applying plus or minus sign corresponds to the correlator for the quasi Sivers function or the quasi Boer-Mulders function, respectively, $\lambda$ is deduced from the on-shell condition of the diquark
\begin{align}
\delta((P-k)^2-M_X^2)&=\delta((P_0-k_0)^2-\bm{k}_T^2-(P_z-k_z)^2-M_X^2)\nonumber \\
&=\frac{1}{2(P_0-k_0)}\delta(P_0-k_0-\lambda)
\label{eq.19},
\end{align}
and has the expression
\begin{align}
\lambda=\sqrt{(1-x)^2P_z^2+\bm{k}_T^2+M_X^2}=(1-x)P_z\rho_X,
\end{align}
with
\begin{align}
\rho_X=\sqrt{1+\frac{\bm{k}^2_T+M_X^2}{(1-x)^2P_z^2}}.
\end{align}

Combining the definitions of the quasi-functions in Eqs.~(\ref{eq.quasi-siv}) and (\ref{eq.quasi-bm}) with the correlator (\ref{eq.14}), it is not difficult to obtain the expressions for the quasi-TMDs
\begin{align}
\tilde{f}_{1T}^{\perp (s)}(x,\bm{k}_T^2,P_z)&=-\frac{g_s(k^2)e_c^2}{4}\frac{1}{(2\pi)^3}\frac{ M}{2\lambda}\frac{2\textrm{Im}\mathcal{I}_1^s}{k^2-m^2},\\
\tilde{h}_1^{\perp (s)}(x,\bm{k}_T^2,P_z)&=-\frac{g_s(k^2)e_c^2}{4}\frac{1}{(2\pi)^3}\frac{ M}{2\lambda}\frac{2\textrm{Im}\mathcal{I}_1^{\prime s}}{k^2-m^2},\\
\tilde{f}_{1T}^{\perp (a)}(x,\bm{k}_T^2,P_z)&=\frac{g_a(k^2)e_c^2}{4}\frac{1}{(2\pi)^3}\frac{ M}{4\lambda}\frac{2\textrm{Im}\mathcal{I}_1^a}{k^2-m^2},\\
\tilde{h}_1^{\perp q(a)}(x,\bm{k}_T^2,P_z)&=\frac{g_a(k^2)e_c^2}{4}\frac{1}{(2\pi)^3}\frac{M} {4\lambda}\frac{2\textrm{Im}\mathcal{I}_1^{\prime a}}{k^2-m^2},
\end{align}
which are the contributions from the scalar diquark and the axial-vector diquark, respectively,
and
\begin{align}
(\epsilon_{T}^{ij}\bm{k}_{Ti}\bm{S}_{Tj})\mathcal{I}_1^s=&\int \frac{d^4l}{(2\pi)^4} g_s((k-l)^2)\frac{  \textrm{Tr}[(\slashed{k}-\slashed{l}-m)(\slashed{P}+M)\gamma_5 \slashed{S}(\slashed{k}+m) (2P-2k+l)_\rho n_z^\rho \gamma_z] }{(D_1+i\epsilon) (D_2+i\epsilon) (D_3+i\epsilon) (D_4+i\epsilon)}
\label{eq:i1s},\\
-(\epsilon_{T}^{ij}\bm{k}_{Tj})\mathcal{I}_1^{\prime s}=&\int \frac{d^4l}{(2\pi)^4}g_s((k-l)^2) \frac{\textrm{Tr}[(\slashed{k}-\slashed{l}-m)(\slashed{P}+M)(\slashed{k}+m) (2P-2k+l)_\rho n_z^\rho i\sigma_{iz}\gamma_5]}{(D_1+i\epsilon) (D_2+i\epsilon) (D_3+i\epsilon) (D_4+i\epsilon)}
\label{eq:i1sp},\\
(\epsilon_{T}^{ij}\bm{k}_{Ti}\bm{S}_{Tj})\mathcal{I}_1^a=&\int \frac{d^4l}{(2\pi)^4} g_a((k-l)^2) \frac{\textrm{Tr}[(\slashed{k}-\slashed{l}+m)\gamma^\mu \gamma_5 (\slashed{P}+M)\gamma_5 \slashed{S} \gamma^\alpha\gamma_5(\slashed{k}+m) d_{\mu\nu}d_{\sigma\alpha}\Gamma_{a\rho}^{\nu\sigma} n_z^\rho \gamma_z] }{(D_1+i\epsilon) (D_2+i\epsilon) (D_3+i\epsilon) (D_4+i\epsilon)}
\label{eq:i1a},\\
-(\epsilon_{T}^{ij}\bm{k}_{Tj})\mathcal{I}_1^{\prime a}=&\int \frac{d^4l}{(2\pi)^4} g_a((k-l)^2)  \frac{ \textrm{Tr}\left[(\slashed{k}-\slashed{l}+m)\gamma^\mu \gamma_5 (\slashed{P}+M) \gamma^\alpha\gamma_5(\slashed{k}+m) d_{\mu\nu}d_{\sigma\alpha}\Gamma_{a\rho}^{\nu\sigma} n_z^\rho i\sigma_{iz}\gamma_5\right] }{(D_1+i\epsilon) (D_2+i\epsilon) (D_3+i\epsilon) (D_4+i\epsilon)}
\label{eq:i1ap},
\end{align}
with
\begin{align}
&D_1=l^2,~~~~ D_3 = (k-l)^2-m^2, \nonumber\\
&D_2=-l^z,~~~~ D_4=((P-k+l)^2-M_s^2. \nonumber
\end{align}
Similarly, we use the residue theorem and pick up the residues of the diquark propagator and eikonal propagator
\begin{align}
\frac{1}{-l^z+i\epsilon} \rightarrow -2\pi i \delta(l^z), \qquad \frac{1}{(P-k+l)^2-M_X^2+i\epsilon} \rightarrow -2\pi i\delta((P-k+l)^2-M_X^2),
\end{align}
where the second delta function provides two solutions
\begin{align}
\delta((P-k-l)^2-M_X^2)={1\over 2\lambda_0}\left(\delta(l_0+\lambda + \lambda_0)+\delta(l_0+\lambda - \lambda_0)\right),
\end{align}
with $\lambda_0=\sqrt{\lambda^2+\bm{l}_T^2-2\bm{l}_T \cdot \bm{k}_T }$.
Here the diquark on-shell condition (\ref{eq.19}) has also been used. For the sake of completeness, both solutions need to be considered.

After computing the traces and the integrals of $l_0$ and $l_z$ in Eqs.~(\ref{eq:i1s}-\ref{eq:i1ap}), we have
\begin{align}
\tilde{f}_{1T}^{\perp (s)}(x,\bm{k}_T^2,P_z)&=\tilde{h}^{\perp q(s)}_1(x,\bm{k}_T^2,P_z)=\frac{1}{(2\pi)^3}\frac{1}{2\rho_s}\frac{e_c^2g_s^2}
{|\bm{k}_T^2+L_s^2(\Lambda_s^2)|^2}M(1-x)^4 \int  \frac{d^2\bm{l}_T}{(2\pi)^2} \frac{1}{\lambda_0}\displaybreak[0]\nonumber \\
&\times \Bigg\{ \frac{(Mk_0+mP_0)\frac{\bm{l}_T\cdot\bm{k}_T}{\bm{k}_T\cdot\bm{k}_T}-M(-\lambda+ \lambda_0)}
{[(-\lambda + \lambda_0)^2-\bm{l}_T^2][(\bm{l}_T-\bm{k}_T)^2+L_s^2(\Lambda_s^2)]^2}
+\frac{(Mk_0+mP_0)\frac{\bm{l}_T\cdot\bm{k}_T}{\bm{k}_T\cdot\bm{k}_T}-M(-\lambda- \lambda_0)}
{[(-\lambda - \lambda_0)^2-\bm{l}_T^2][(\bm{l}_T-\bm{k}_T)^2+L_s^2(\Lambda_s^2)]^2} \Bigg\} ,\displaybreak[0]\label{eq:f1ts}\\
\tilde{f}_{1T}^{\perp (a)}(x,\bm{k}_T^2,P_z)&=\frac{1}{(2\pi)^3}\frac{1}{4\rho_a}\frac{e_c^2g_a^2}
{|\bm{k}_T^2+L_a^2(\Lambda_a^2)|^2}M(1-x)^3 \int  \frac{d^2\bm{l}_T}{(2\pi)^2} \frac{1}{\lambda_0}\displaybreak[0] \\
&\times \Bigg\{\frac{[(m-3M)k_0-2(m-M)xP_0]\frac{\bm{l}_T\cdot\bm{k}_T}{\bm{k}_T\cdot\bm{k}_T}+(3M-m)(-\lambda+ \lambda_0)}
{[(-\lambda + \lambda_0)^2-\bm{l}_T^2][(\bm{l}_T-\bm{k}_T)^2+L_a^2(\Lambda_a^2)]^2}\displaybreak[0]\nonumber \\
&+\frac{[(m-3M)k_0-2(m-M)xP_0]\frac{\bm{l}_T\cdot\bm{k}_T}{\bm{k}_T\cdot\bm{k}_T}+(3M-m)(-\lambda- \lambda_0)}
{[(-\lambda - \lambda_0)^2-\bm{l}_T^2][(\bm{l}_T-\bm{k}_T)^2+L_a^2(\Lambda_a^2)]^2} \Bigg\},\displaybreak[0]\label{eq:f1ta}\\
\tilde{h}^{\perp (a)}_1(x,\bm{k}_T^2,P_z)&=\frac{1}{(2\pi)^3}\frac{1}{4\rho_a}
\frac{e_c^2g_a^2}{|\bm{k}_T^2+L_a^2(\Lambda_a^2)|^2}M(1-x)^3 \int  \frac{d^2\bm{l}_T}{(2\pi)^2} \frac{1}{\lambda_0}\displaybreak[0]\nonumber \\
&\times \Bigg\{\frac{[-(m+M(2x-3))k_0+2mP_0]\frac{\bm{l}_T\cdot\bm{k}_T}
{\bm{k}_T\cdot\bm{k}_T}+(m+M(2x-3))(-\lambda+ \lambda_0)}
{[(-\lambda + \lambda_0)^2-\bm{l}_T^2][(\bm{l}_T-\bm{k}_T)^2+L_a^2(\Lambda_a^2)]^2}\displaybreak[0]\nonumber \\
&+\frac{[-(m+M(2x-3))k_0+2mP_0]\frac{\bm{l}_T\cdot\bm{k}_T}{\bm{k}_T\cdot\bm{k}_T}+(m+M(2x-3))(-\lambda- \lambda_0)}
{[(-\lambda - \lambda_0)^2-\bm{l}_T^2][(\bm{l}_T-\bm{k}_T)^2+L_a^2(\Lambda_a^2)]^2} \Bigg\} \label{eq:h1ta}.
\end{align}
The integration over $\bm l_T$ can be performed numerically.
With the above results, we can obtain the quasi Sivers function and the quasi Boer-Mulders function of the $u$ and $d$ quarks using the relation similar to Eq.~(\ref{eq:ud})
\begin{align}
\tilde{f}^{ u} = {3\over 2} \tilde{f}^{ (s)} + {1\over 2} \tilde{f}^{ (a)}, ~~~~~ \tilde{f}^d = \tilde{f}^{ (a)} . \label{eq:ud2}
\end{align}
We can also provide the first transverse-moment of these quasi-functions:
\begin{align}
\tilde{f}_{1T}^{\perp (1)}(x,P_z)&=\int d^2\bm{k}_T \frac{\bm{k}_T^2}{2M^2}\tilde{f}_{1T}^{\perp q}(x,\bm{k}_T^2,P_z),\label{eq:1st-trans}\\
\tilde{h}_1^{\perp (1)}(x,P_z)&=\int d^2\bm{k}_T \frac{\bm{k}_T^2}{2M^2}\tilde{h}_1^{\perp q}(x,\bm{k}_T^2,P_z).
\end{align}

Finally, we find that the T-odd TMDs reduce to the standard TMDs in the limit $P_z \rightarrow \infty$,
\begin{align}
\tilde{f}_{1T}^{\perp q}(x,\bm{k}_T^2,P_z \rightarrow \infty)&=f_{1T}^{\perp q}(x,\bm{k}_T^2),\\
\tilde{h}^{\perp q}_1(x,\bm{k}_T^2,P_z \rightarrow \infty)&=h^{\perp q}_1(x,\bm{k}_T^2).
\end{align}
This extends the observation that the quasi-PDFs should reduce to the standard PDFs defined in terms of the light-cone correlation functions~\cite{Ji:2013dva,Ma:2014jla} at $P_z \rightarrow \infty$.

\section{Numerical result} \label{Sec:4}

In this section, we will numerically compute the T-odd quasi-TMDs of $u$ and $d$ quarks to study the $P_z$ dependence of these distributions, and compare them with the standard distributions.
To do this we need to assign the values of the parameters in Eqs.~(\ref{eq:f1ts}-\ref{eq:h1ta}). Here we adopt the choices in the original works~\cite{Jakob:1997wg,Bacchetta:2003rz}
\begin{align}
&m= 0.36\ \textrm{GeV},~~~~~~~~~ g_s^2=6.525, ~~~~~~~~~~~~g_a^2 =28.716,\\
&\Lambda_{s/a}=0.5 \ \textrm{GeV},~~~~~~~M_s=0.6\ \textrm{GeV}, ~~~~~~~~M_a=0.8 \ \textrm{GeV}.
\end{align}
The factors $g_s$ and $g_a$ are determined from the normalization condition of the unpolarized distributions
\begin{align}
\pi \int^1_0dx\int^\infty_0d\bm{k}_T^2 {f}_1^{(s)}(x,\bm k_T^2)=1,\qquad
\pi \int^1_0dx\int^\infty_0d\bm{k}_T^2 f_1^{ (a)}(x,\bm k_T^2)=1,
\end{align}
which consequently normalizes $f_1^u$ to 2 and $f_1^d$ to 1.

In order to replace the Abelian interaction of gluons with the QCD color interaction, we make the following replacement~\cite{Brodsky:2002cx}
\begin{align}
{e_c^2\over 4\pi}\rightarrow  C_F \alpha_s,
\end{align}
where $C_F=4/3$ and we choose $\alpha_s\approx 0.3$ in the calculation.

\begin{figure}
  \centering
  \includegraphics[width=0.4\columnwidth]{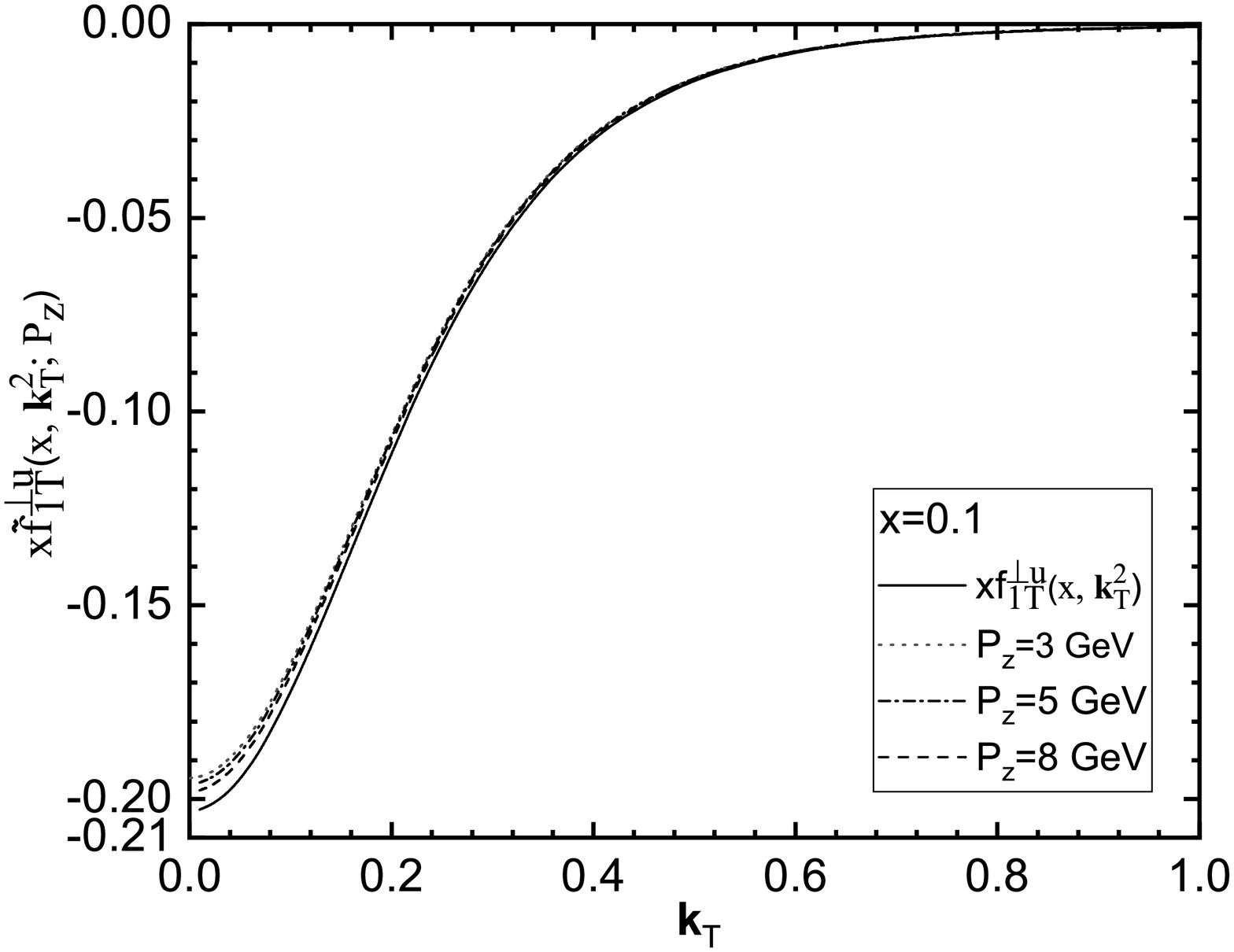}~~~
  \includegraphics[width=0.4\columnwidth]{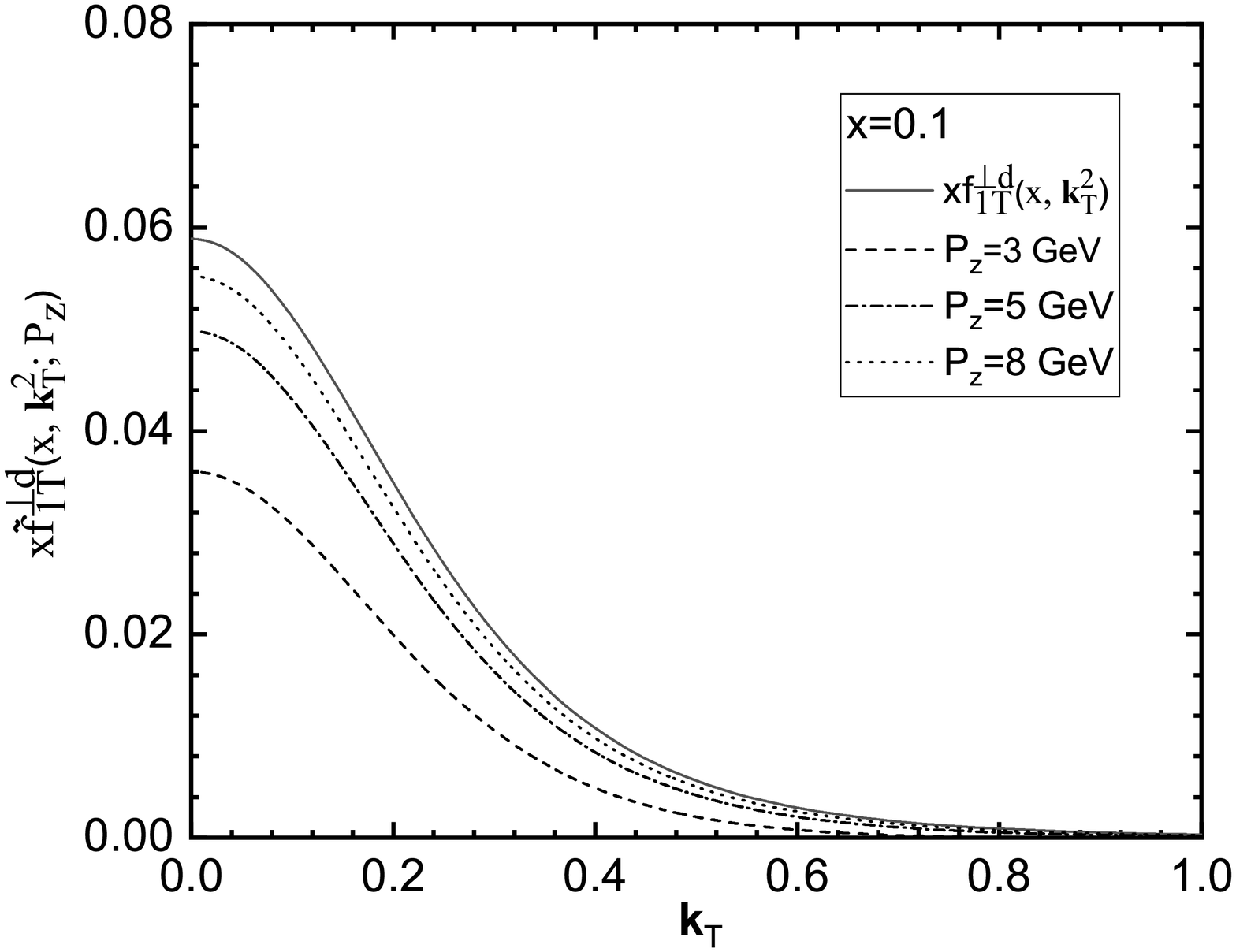}\\
  \includegraphics[width=0.4\columnwidth]{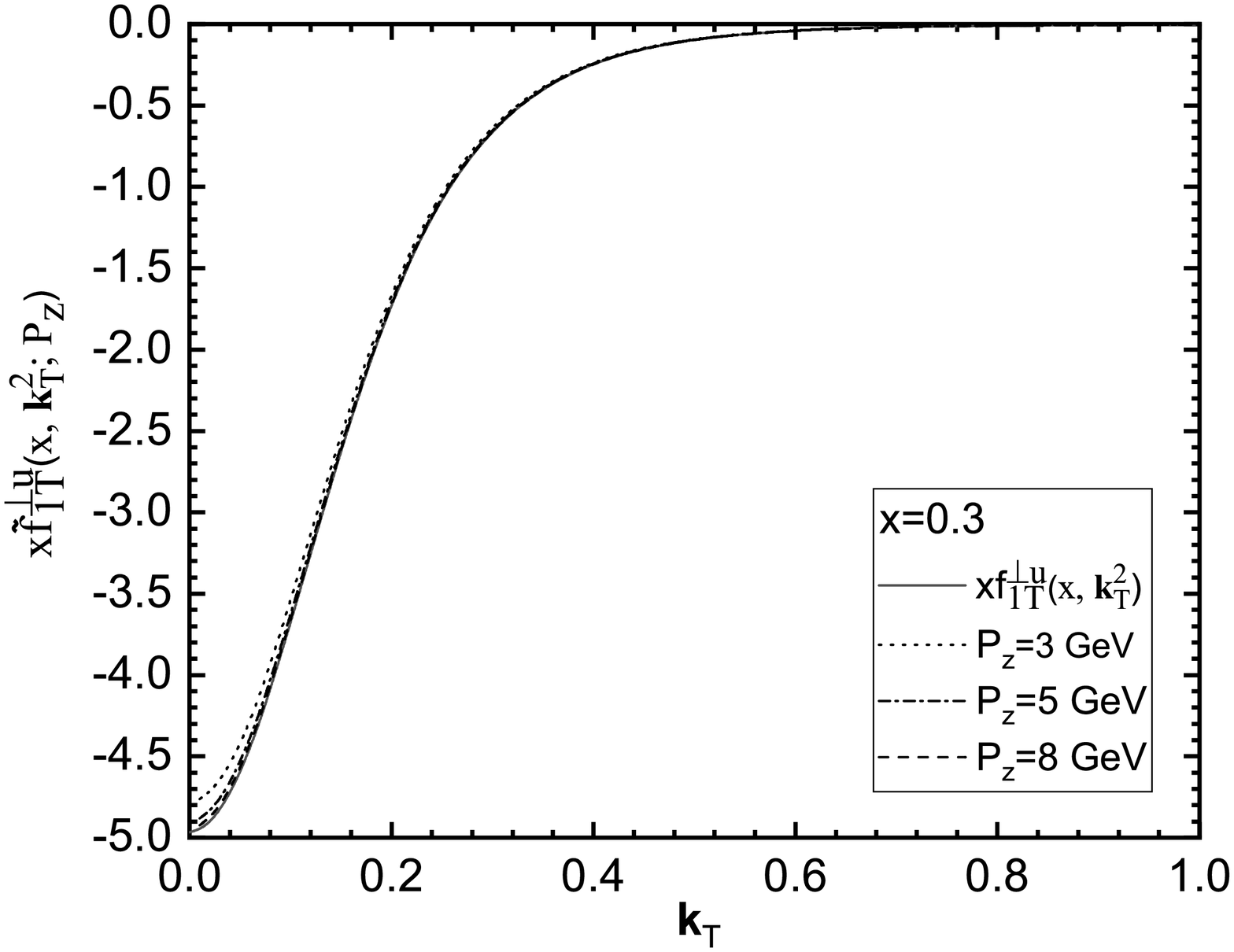}~~~
  \includegraphics[width=0.4\columnwidth]{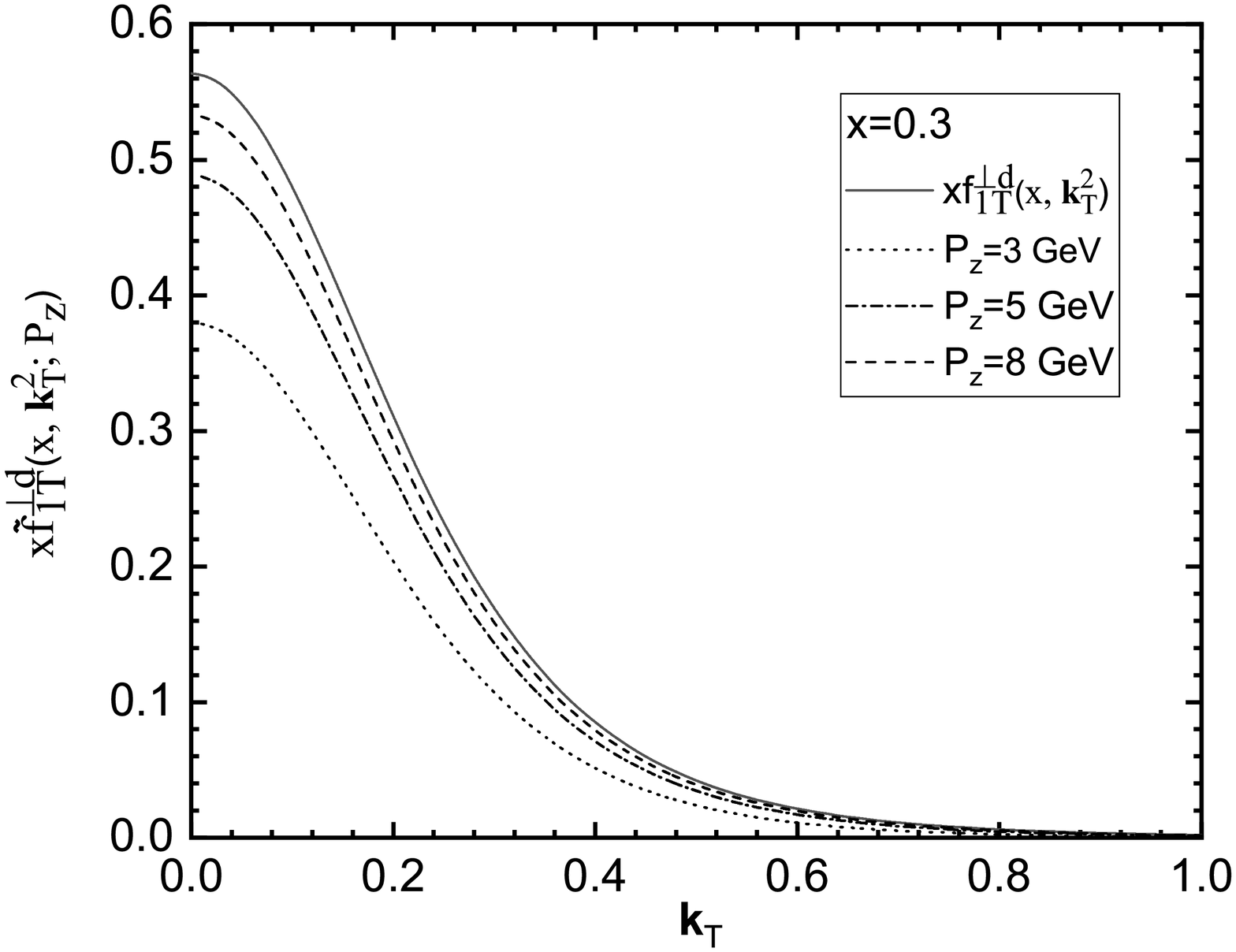}\\
  \includegraphics[width=0.4\columnwidth]{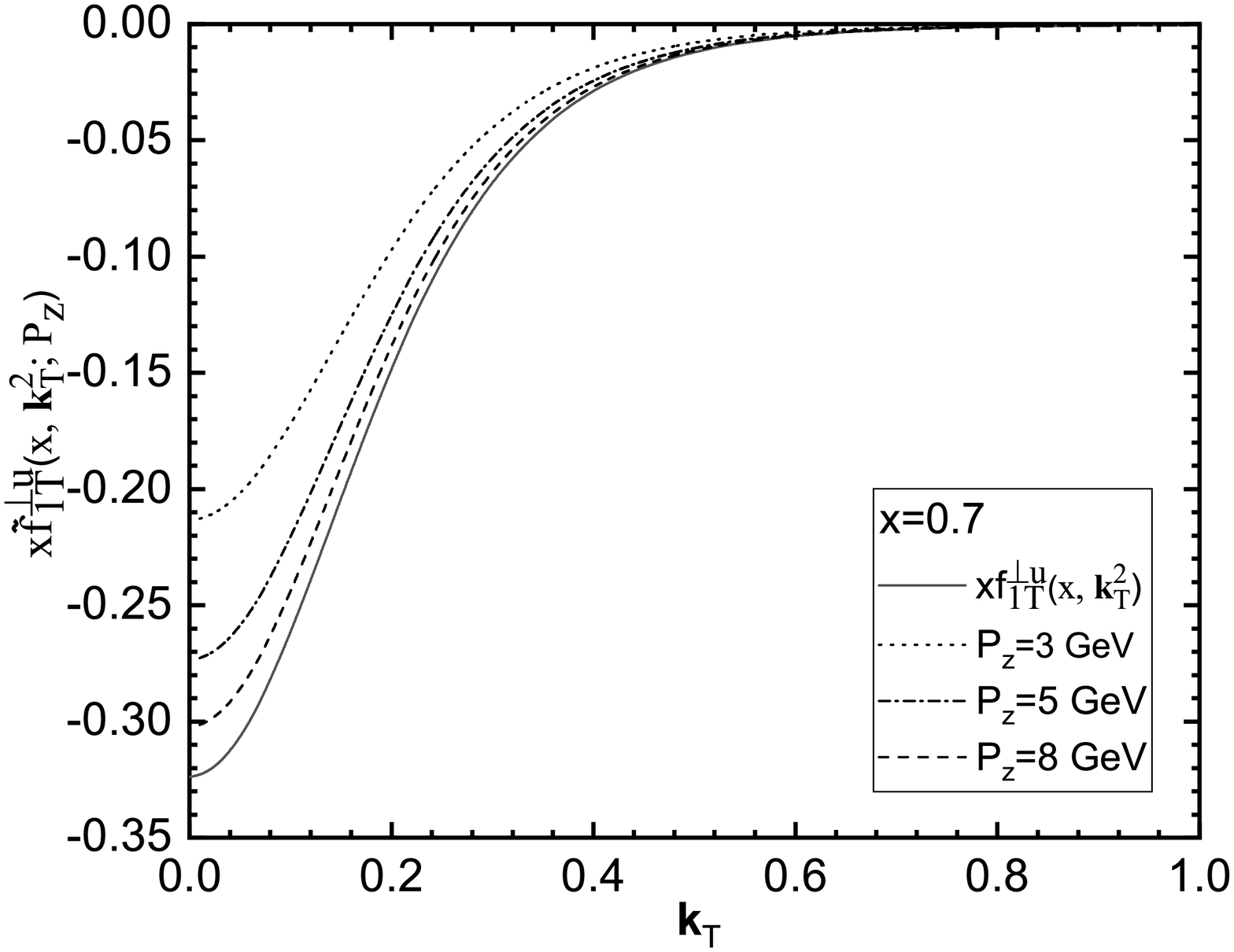}~~~
  \includegraphics[width=0.4\columnwidth]{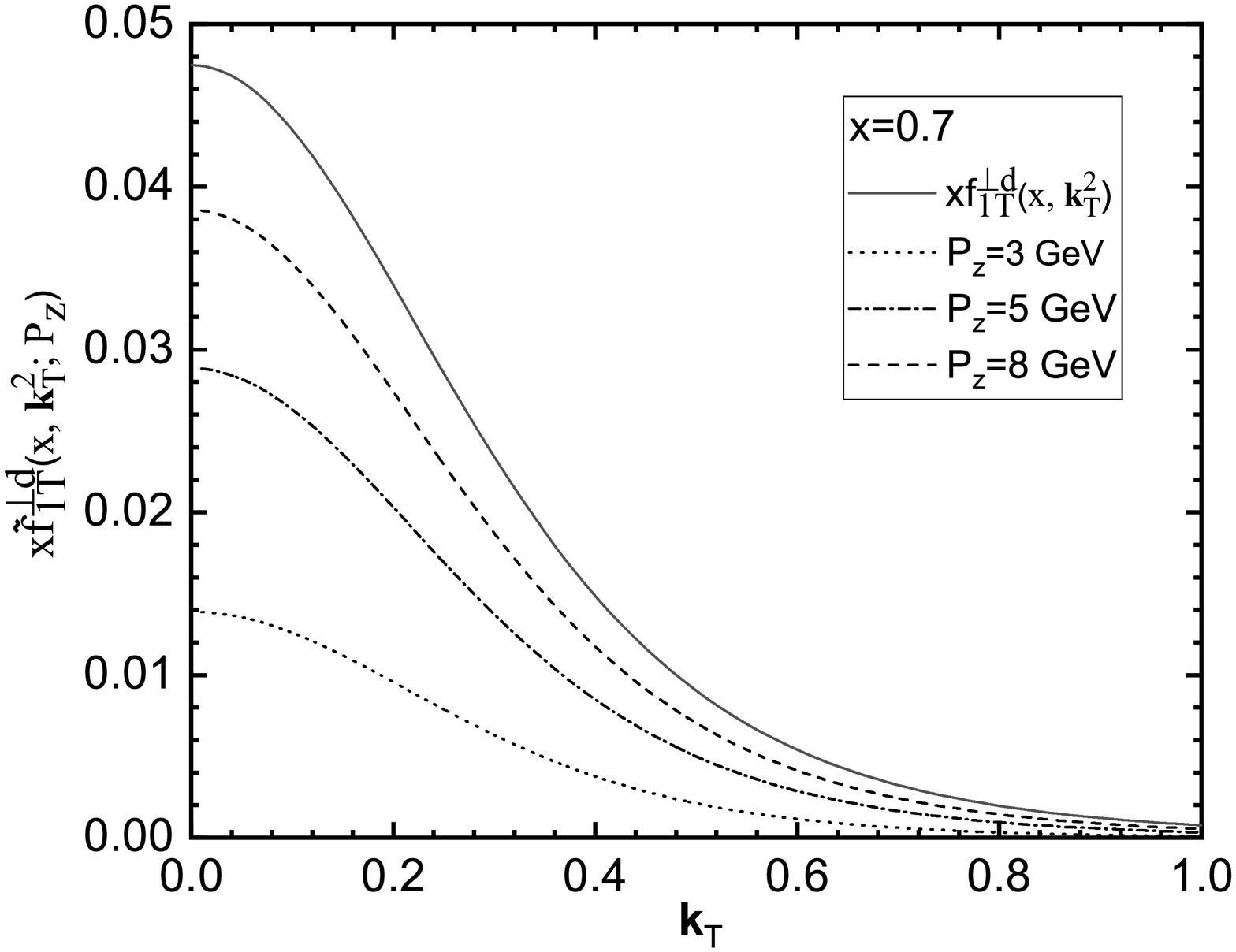}\\
  \caption{The quasi Sivers function $\tilde{f}^{\perp q}_{1T}(x,\bm{k}_T^2;P_z)$ (multiplied by $x$) of the up (left panel) and down (right panel) quarks as a function of $k_T$ in the spectator model. The dotted line, the dotted-dashed line and the dashed line correspond to the results at $P_z=3$ GeV, 5 GeV and 8 GeV, respectively.}
  \label{fig:quasi-siv-kt}
\end{figure}

\begin{figure}
  \centering
  \includegraphics[width=0.4\columnwidth]{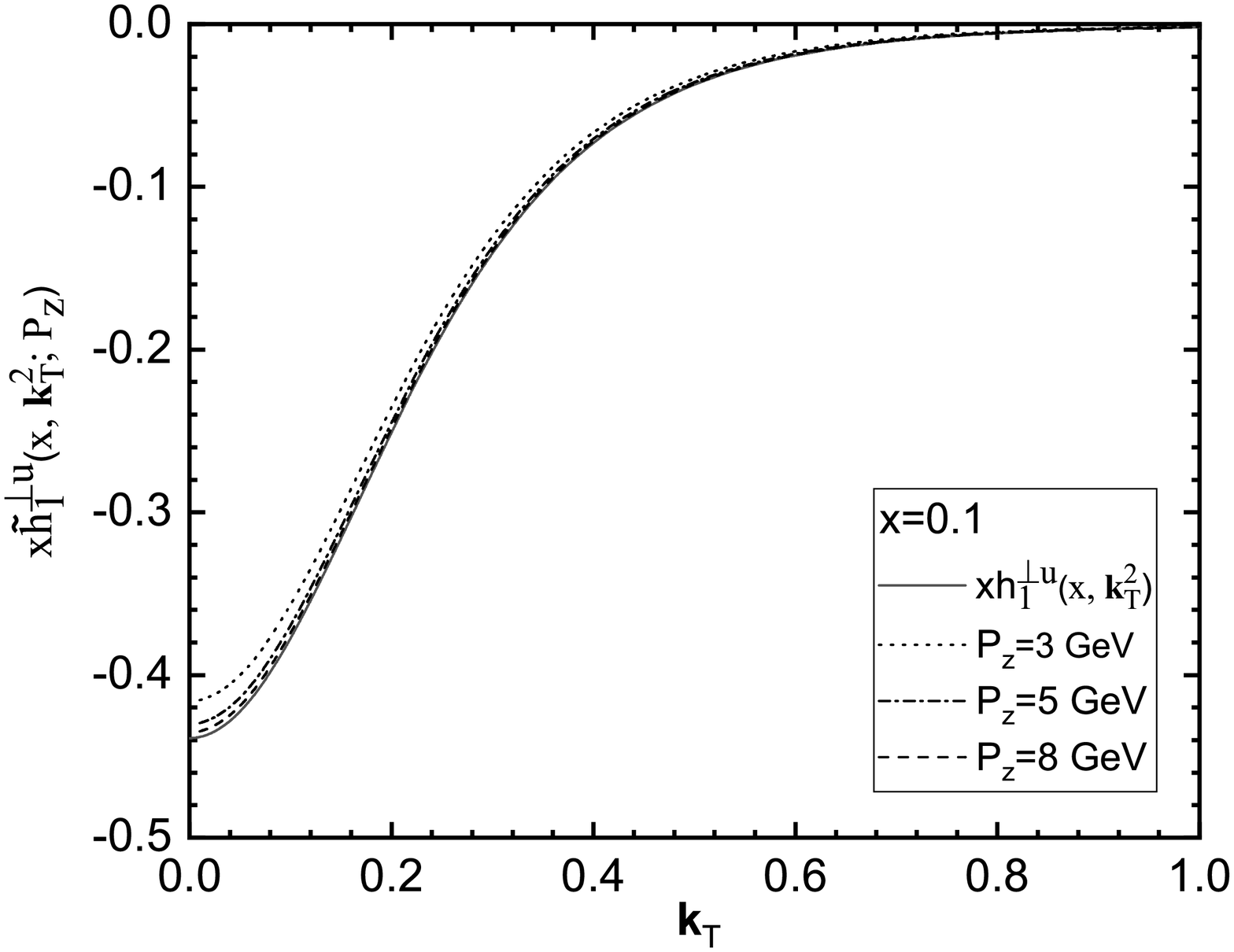}~~~
  \includegraphics[width=0.4\columnwidth]{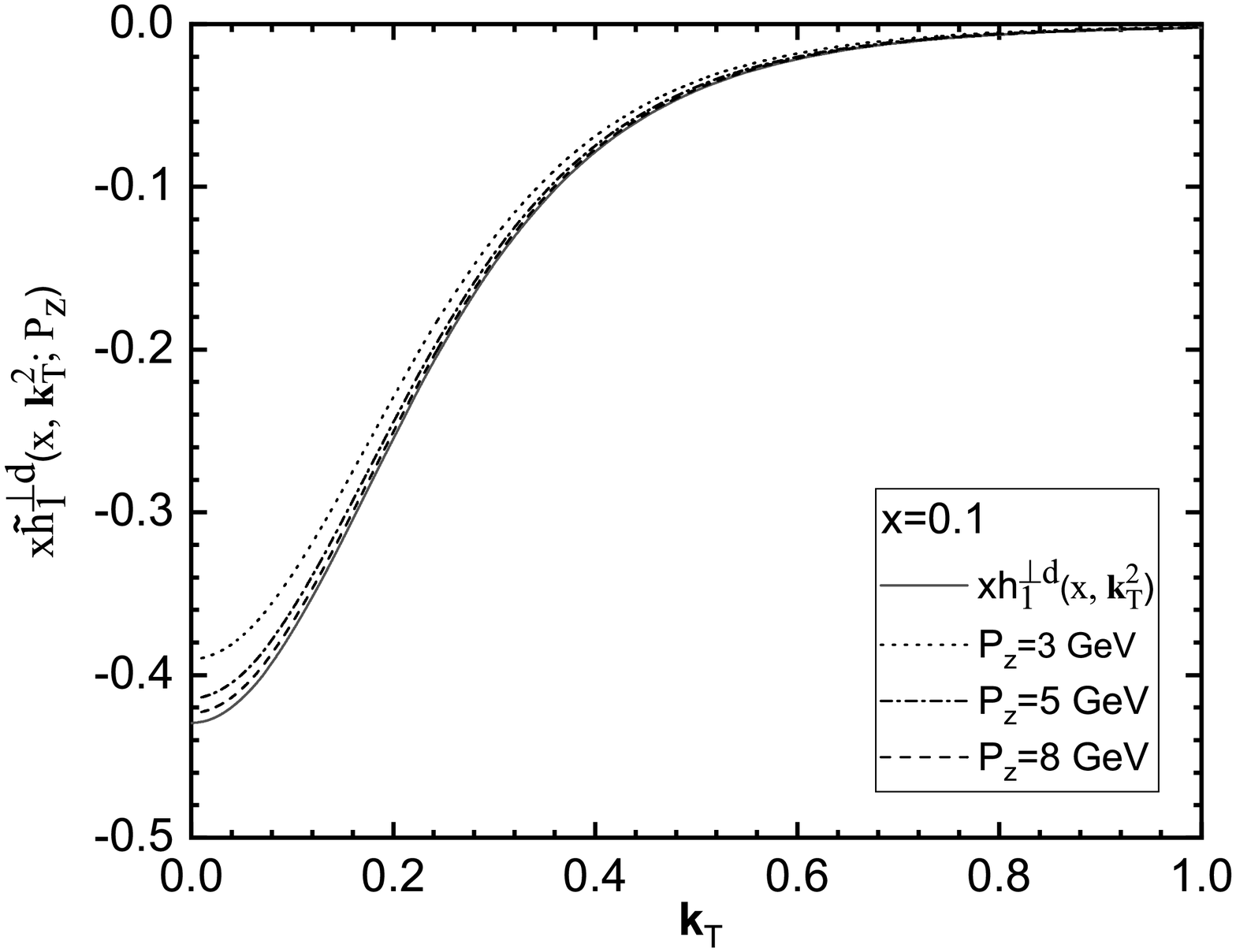}\\
  \includegraphics[width=0.4\columnwidth]{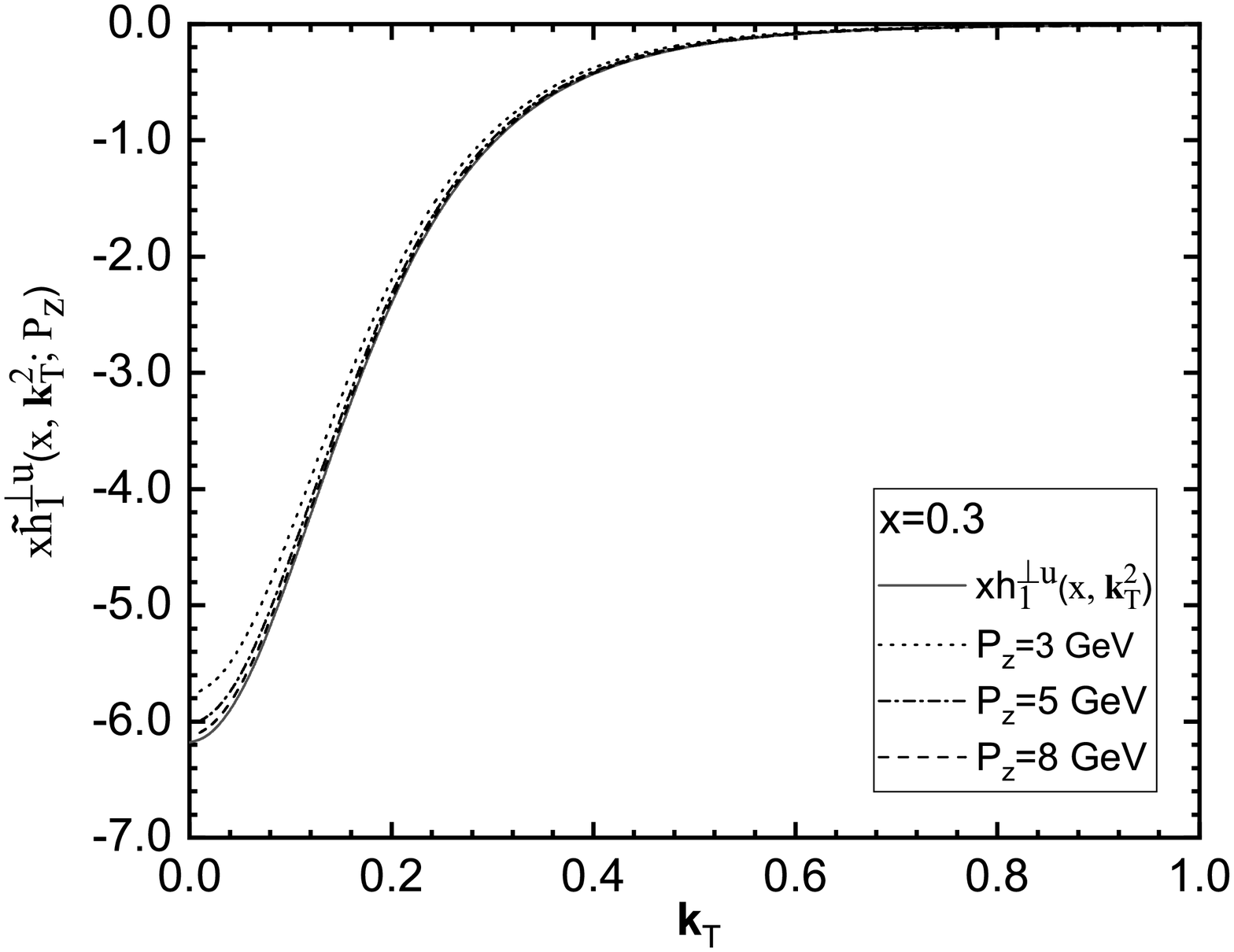}~~~
  \includegraphics[width=0.4\columnwidth]{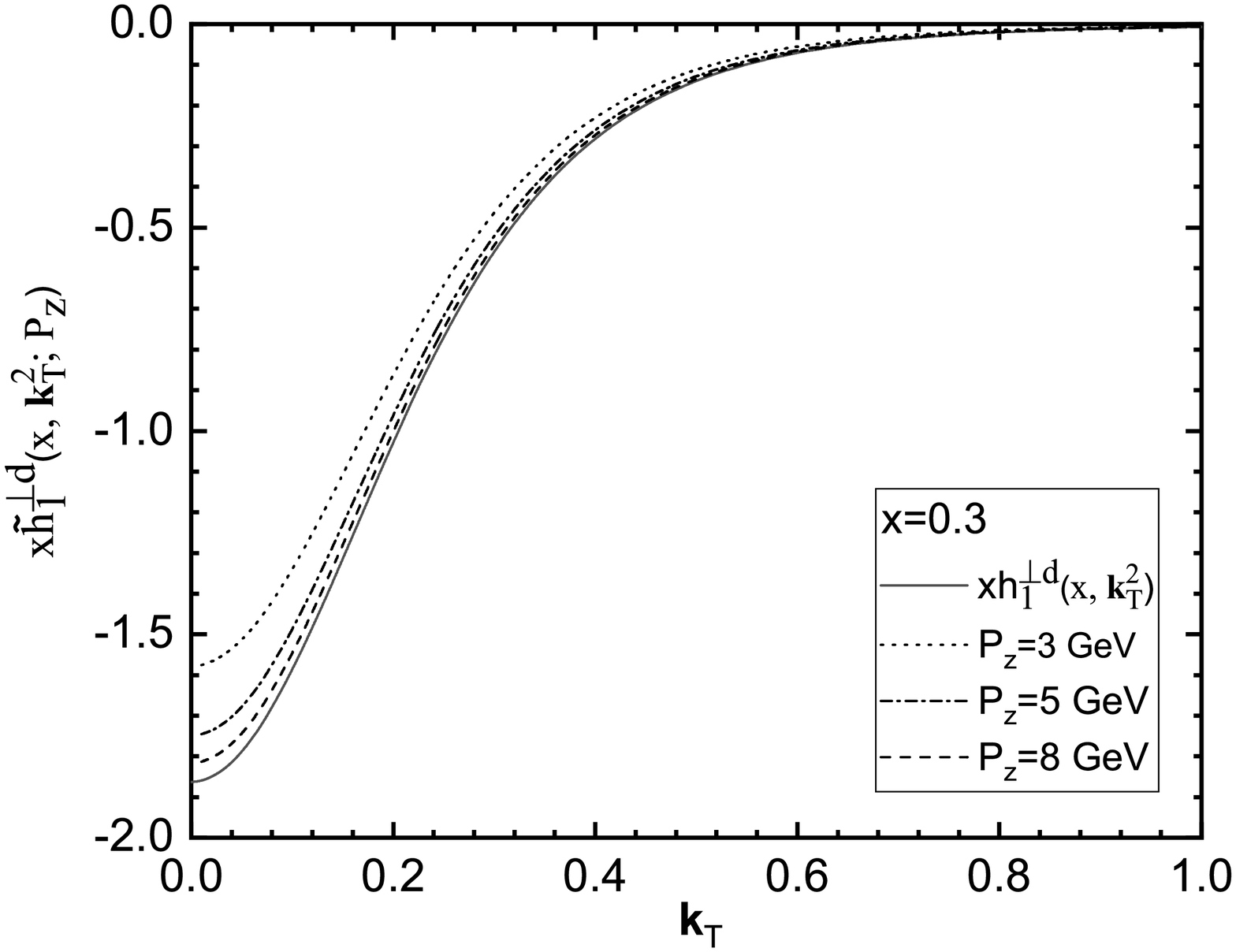}\\
  \includegraphics[width=0.4\columnwidth]{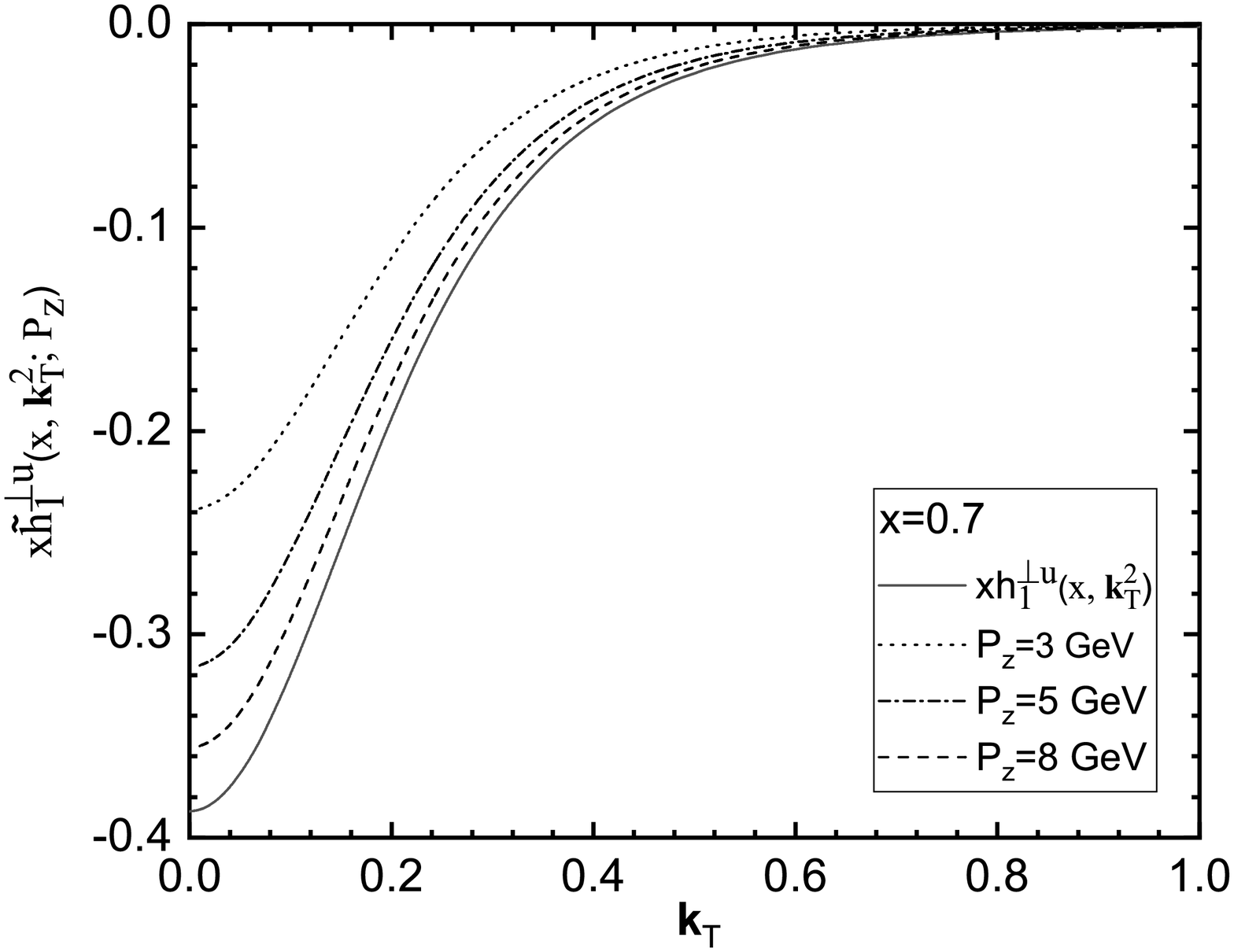}~~~
  \includegraphics[width=0.4\columnwidth]{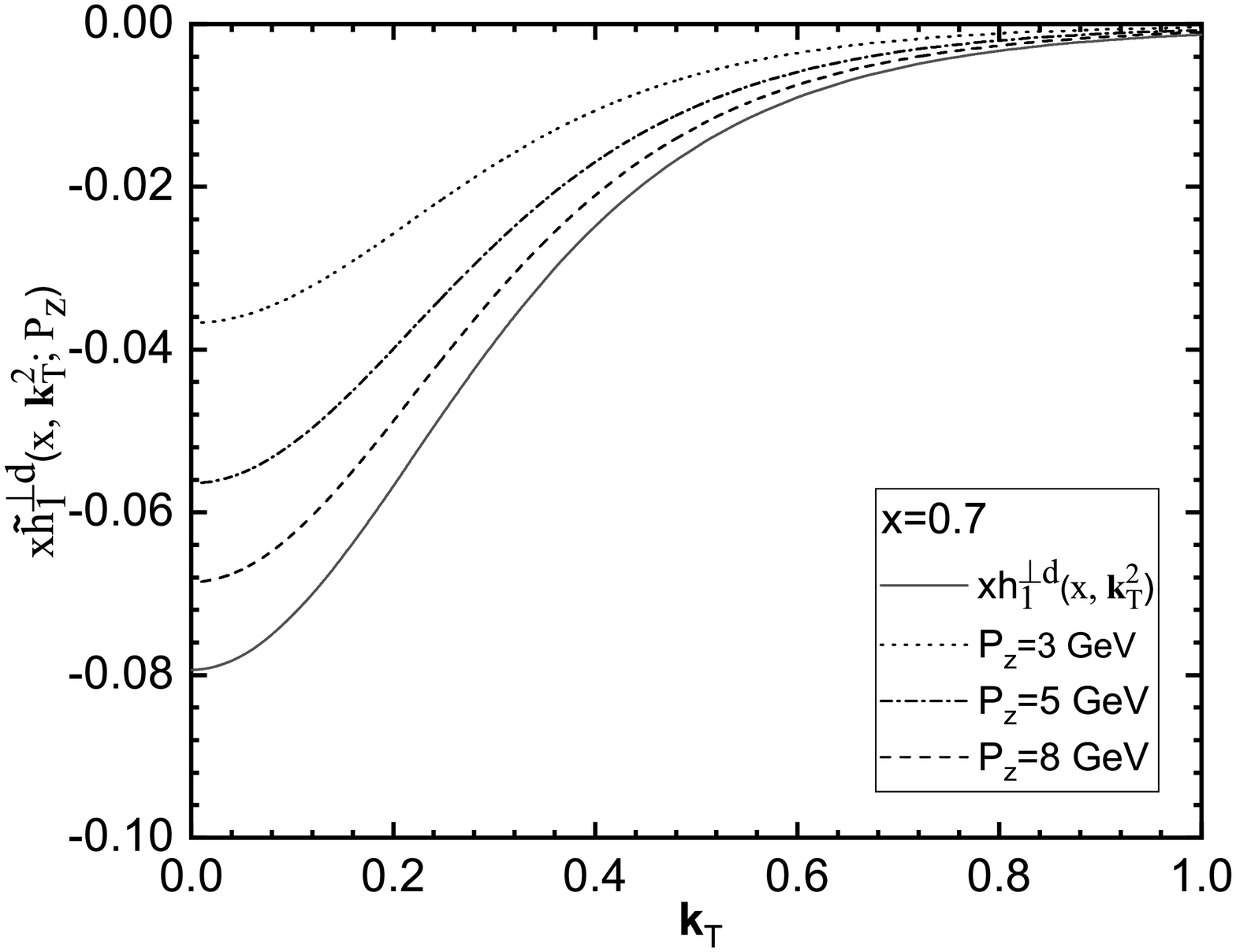}
  \caption{The $x\tilde{f}^{\perp (1)u/d}_{1T}(x,P_z,\bm{k}_T^2)$ and $x\tilde{h}^{\perp (1)u/d}_1(x,P_z,\bm{k}_T^2)$ at $P_z =3$ GeV, 5 GeV and 8 GeV in the spectator model. The upper panel, central panel and lower panel show the results at $x=0.1$, 0.3, 0.7, respectively}
  \label{fig:quasi-bm-kt}
\end{figure}

\begin{figure}
  \centering
  \includegraphics[width=0.4\columnwidth]{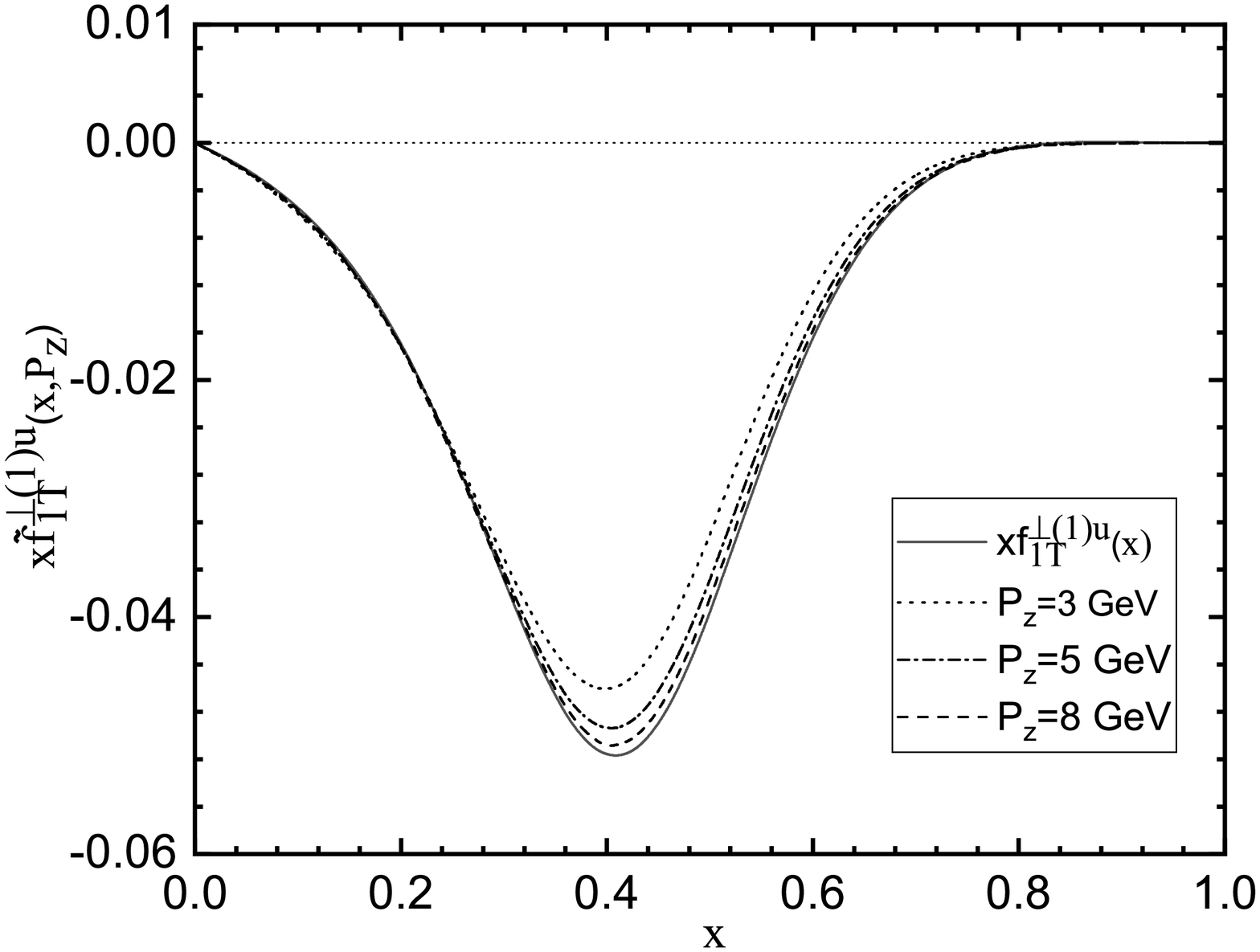}~~~
  \includegraphics[width=0.4\columnwidth]{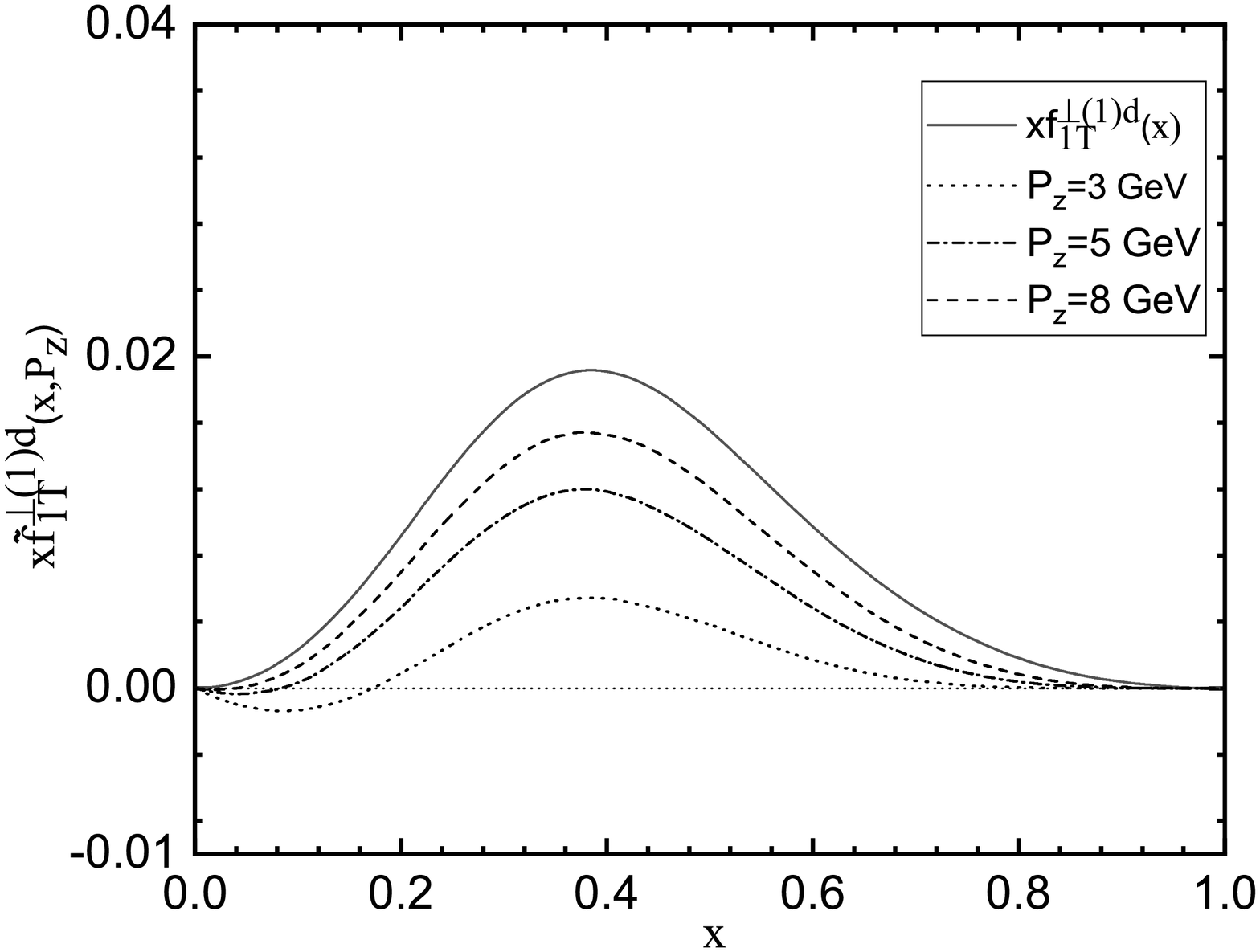}\\
  \includegraphics[width=0.4\columnwidth]{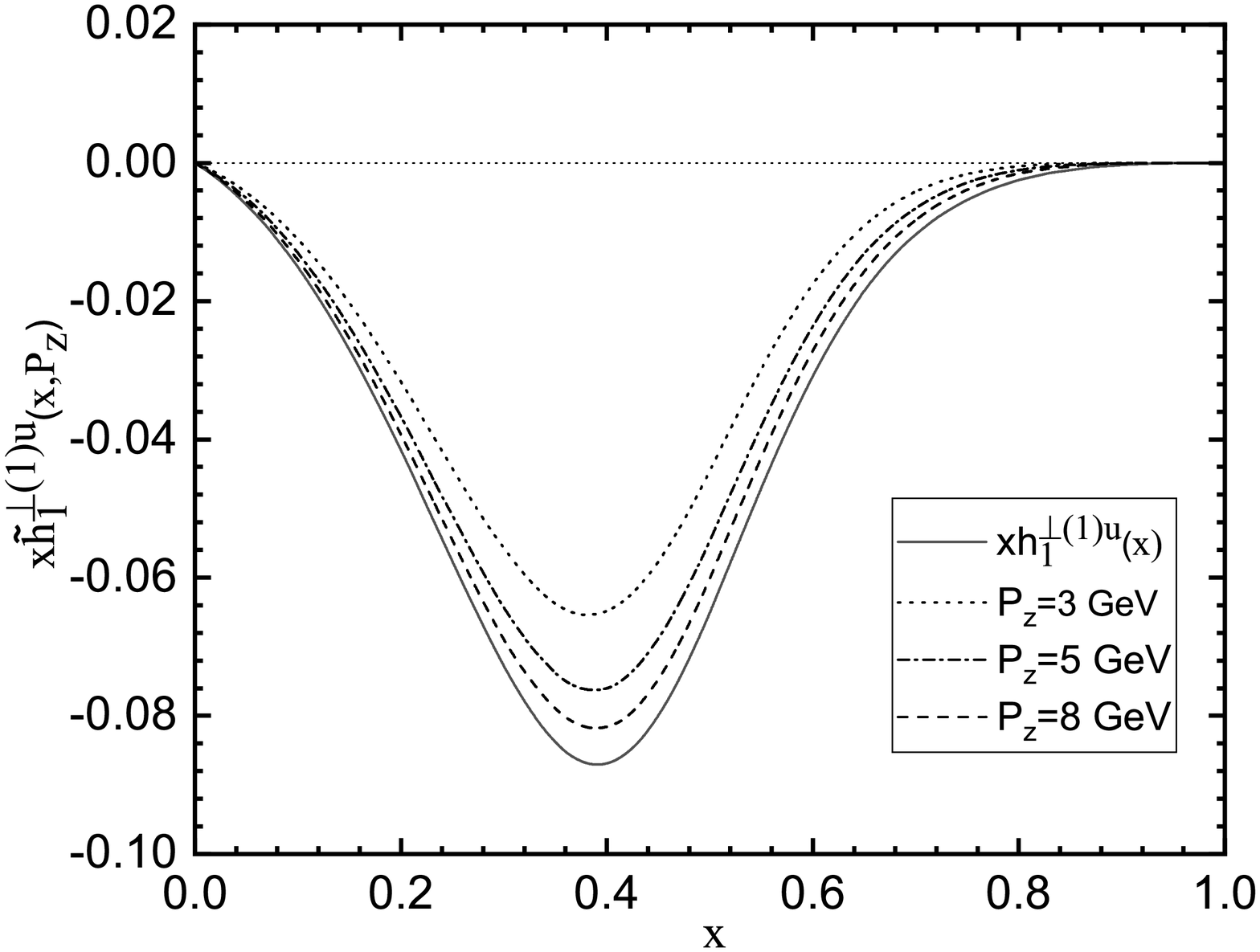}~~~
  \includegraphics[width=0.4\columnwidth]{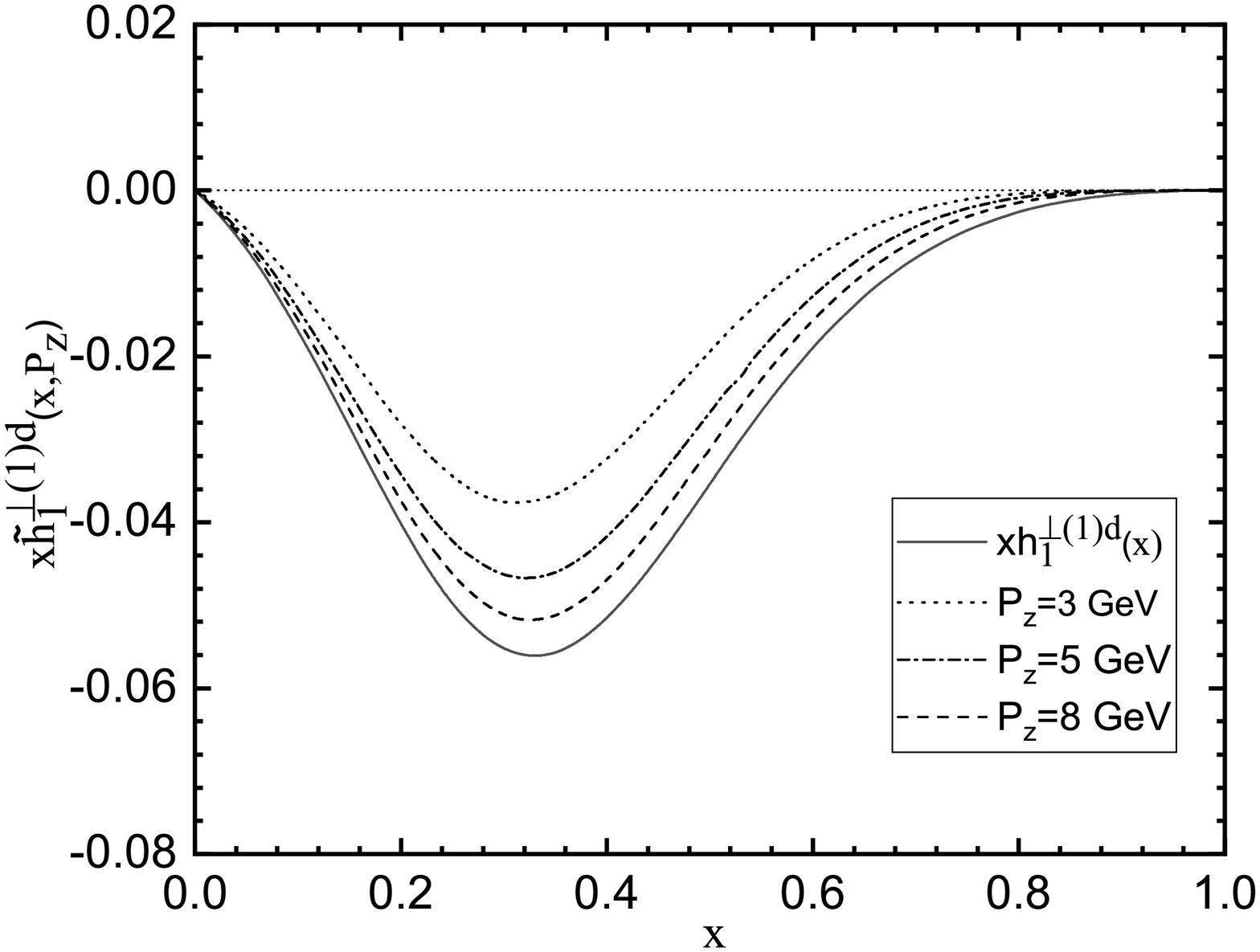}
  \caption{The first $\bm{k}_T$-moment of the quasi Sivers function $x\tilde{f}^{\perp q}_{1T}(x,P_z)$ (upper panel) and that of the quasi Boer-Mulders function $x\tilde{h}^{\perp (1) q}_1(x,P_z)$ (lower panel) as a function of $x$ in the spectator model.
  The dotted, dash-dotted, dashed lines denote the results at $P_z=3$ GeV, 5 GeV and 8 GeV, respectively.
  The solid lines depict the corresponding first $\bm{k}_T$-moment of the standard functions.}
  \label{fig:1st-trans}
\end{figure}

In Fig.~\ref{fig:quasi-siv-kt}, we plot the quasi Sivers function (timed with $x$) $\tilde{f}_{1T}^{\perp}(x,\bm k_T^2; P_z)$ of the up (left panel) and down (right panel) quarks as a function of $k_T=|\bm k_T|$.
The upper panel, central panel and lower panel show the results at $x=0.1$, 0.3, 0.7, respectively.
The dotted line, the dotted-dashed line and the dashed line correspond to the results at $P_z=3$ GeV, 5 GeV and 8 GeV, respectively.
The solid line shows the result of the standard Sivers function $f_{1T}^{\perp}(x,\bm k_T^2)$ for comparison.
The quasi Boer-Mulders function (timed with $x$) $\tilde{h}_{1}^{\perp}(x,\bm k_T^2; P_z)$  is plotted in Fig.~\ref{fig:quasi-bm-kt} in a similar way.
One can find that the sizes of the quasi-TMDs decrease with increasing $k_T$, which is similar to the $k_T$-shape of the standard TMDs.
The results also show that the quasi Sivers function of the up quark is negative, while that of the down quark is positive.
The quasi Boer-Mulders functions of the up and down quarks are both negative.
In all cases the sizes of the T-odd quasi-TMDs are smaller than those of the standard TMDs.
However, as $P_z$ increases, the sizes of the quasi-TMDs increase and converge to the standard TMDs.
Another observation is that the convergence depends on $x$, that is, in the smaller $x$ region, in general the quasi-TMDs converge more quickly as $P_z$ increases.

In the upper panel of Fig.~\ref{fig:1st-trans}, we plot the $x$-dependence of the first transverse-moment of the quasi Sivers function $\tilde{f}_{1T}^{\perp (1)}(x,P_z)$ (timed with $x$) of the $u$ (left panel) and $d$ (right panel) quarks defined in Eq.~(\ref{eq:1st-trans}).
The dotted line, the dotted-dashed line and the dashed line correspond to $P_z=3$ GeV, 5 GeV and 8 GeV, respectively.
The solid line denotes the first transverse-moment of the standard Sivers function ${f}_{1T}^{\perp (1)}(x)$.
Similarly, we plot the $x$-dependence of the first transverse-moment of the quasi Boer-Mulders function $\tilde{h}_{1}^{\perp (1)}(x,P_z)$ (timed with $x$) of the $u$ (left panel) and $d$ (right panel) quarks in the lower panel of Fig.~\ref{fig:1st-trans}.
Again, here the solid line denotes the first transverse-moment of the standard Boer-Mulders function.
We note that our results of $f_{1T}^{\perp\ (1)}(x)$ and $h_{1}^{\perp (1)}(x)$ qualitatively consistent with the phenomenologically extractions in size and sign.
We find that as $P_z$ increases, in general the sizes of $\tilde{f}_{1T}^{\perp (1)}(x,P_z)$ and $\tilde{h}_{1}^{\perp (1)}(x,P_z)$ increase and the shapes of them get close to the corresponding standard distributions.
There are also some exceptions which can be seen in the small $x$ region of the quasi Sivers function,
particularly, for the $d$ quark the quasi function can have a sign opposite to that of the standard function in the small region. Also a node appears at $x\approx 0.2 $ for $\tilde{f}_{1T}^{\perp (1) d}(x,P_z)$ at small $P_z$ region.

\begin{figure}
  \centering
  \includegraphics[width=0.4\columnwidth]{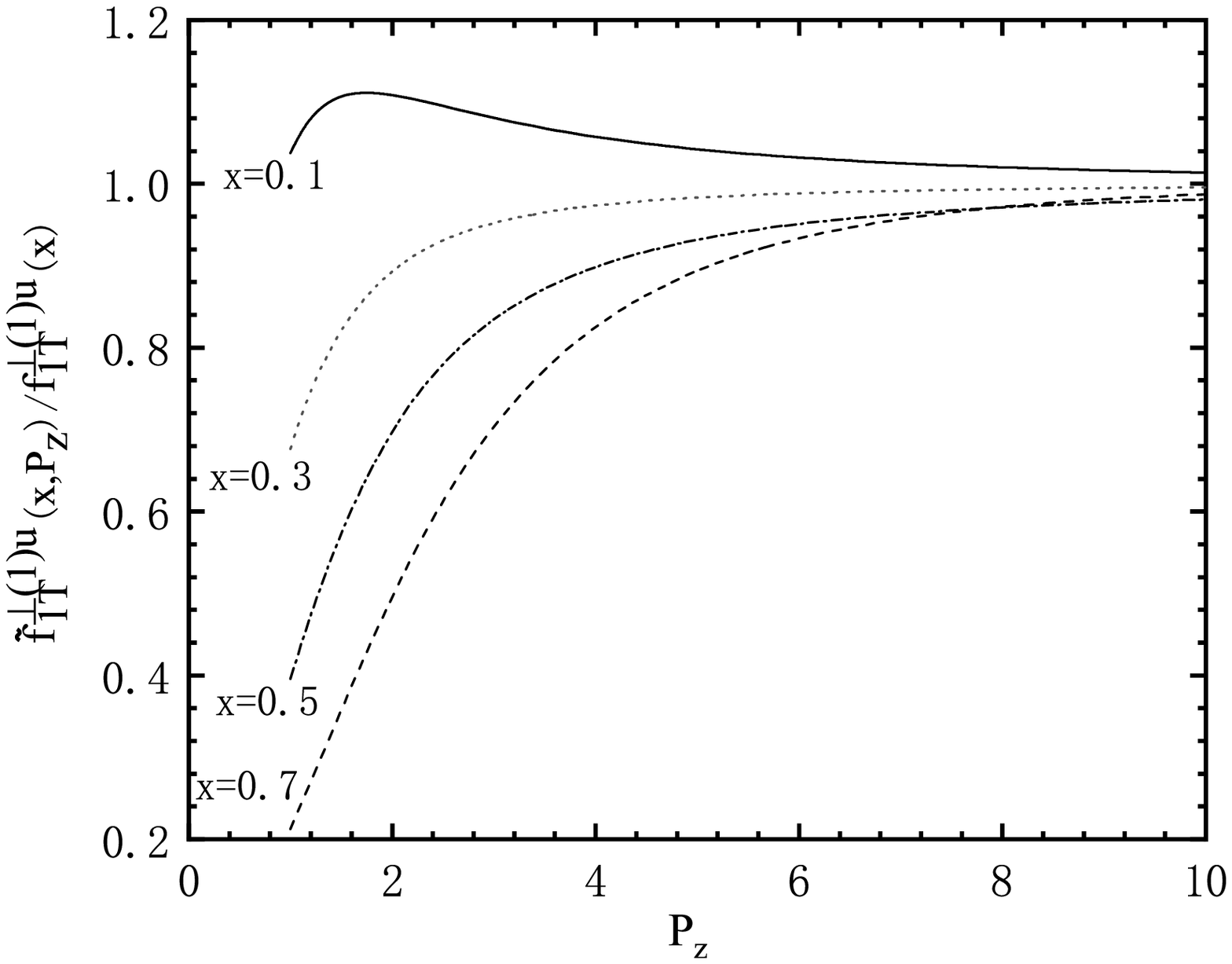}~~~
  \includegraphics[width=0.4\columnwidth]{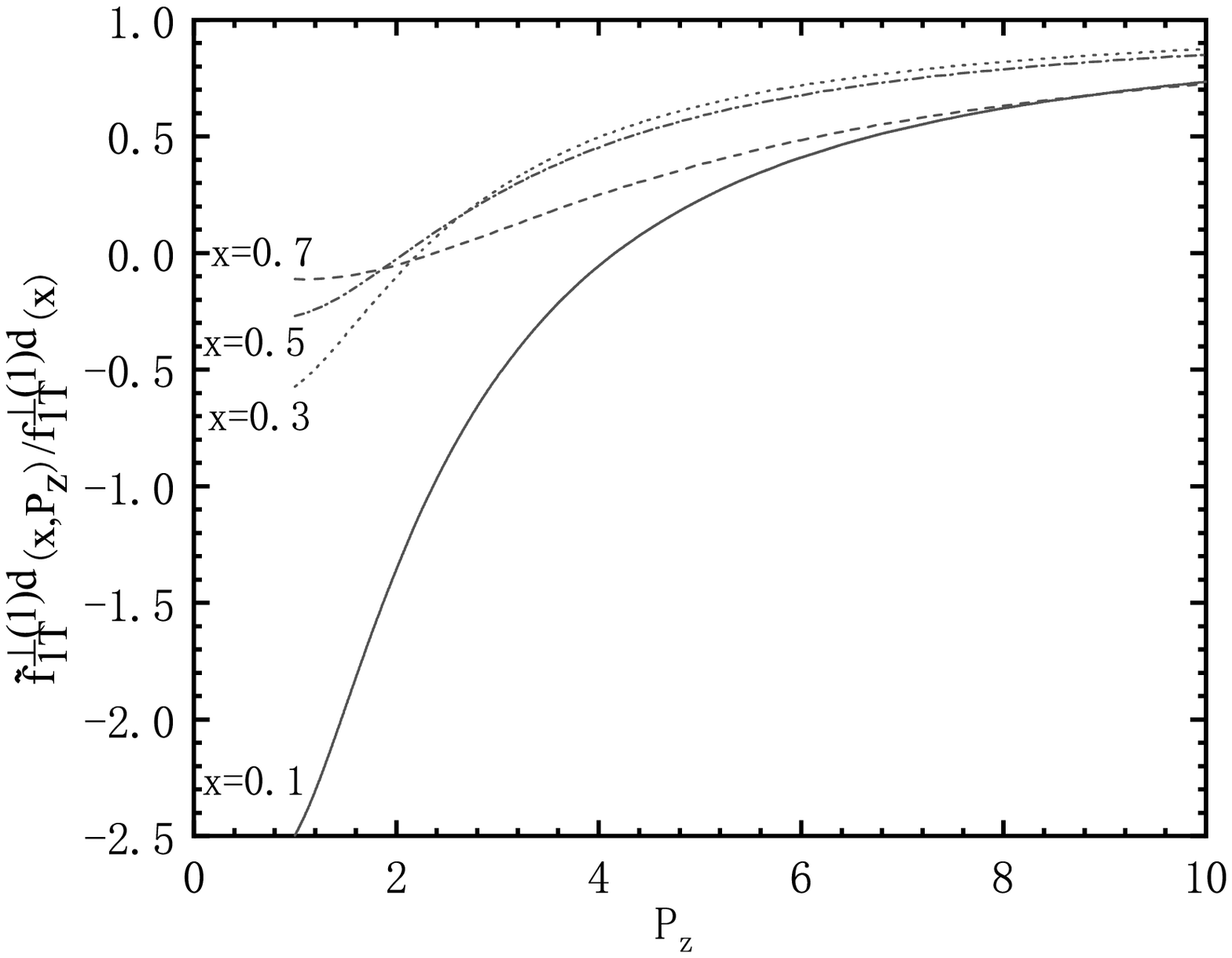}\\
  \includegraphics[width=0.4\columnwidth]{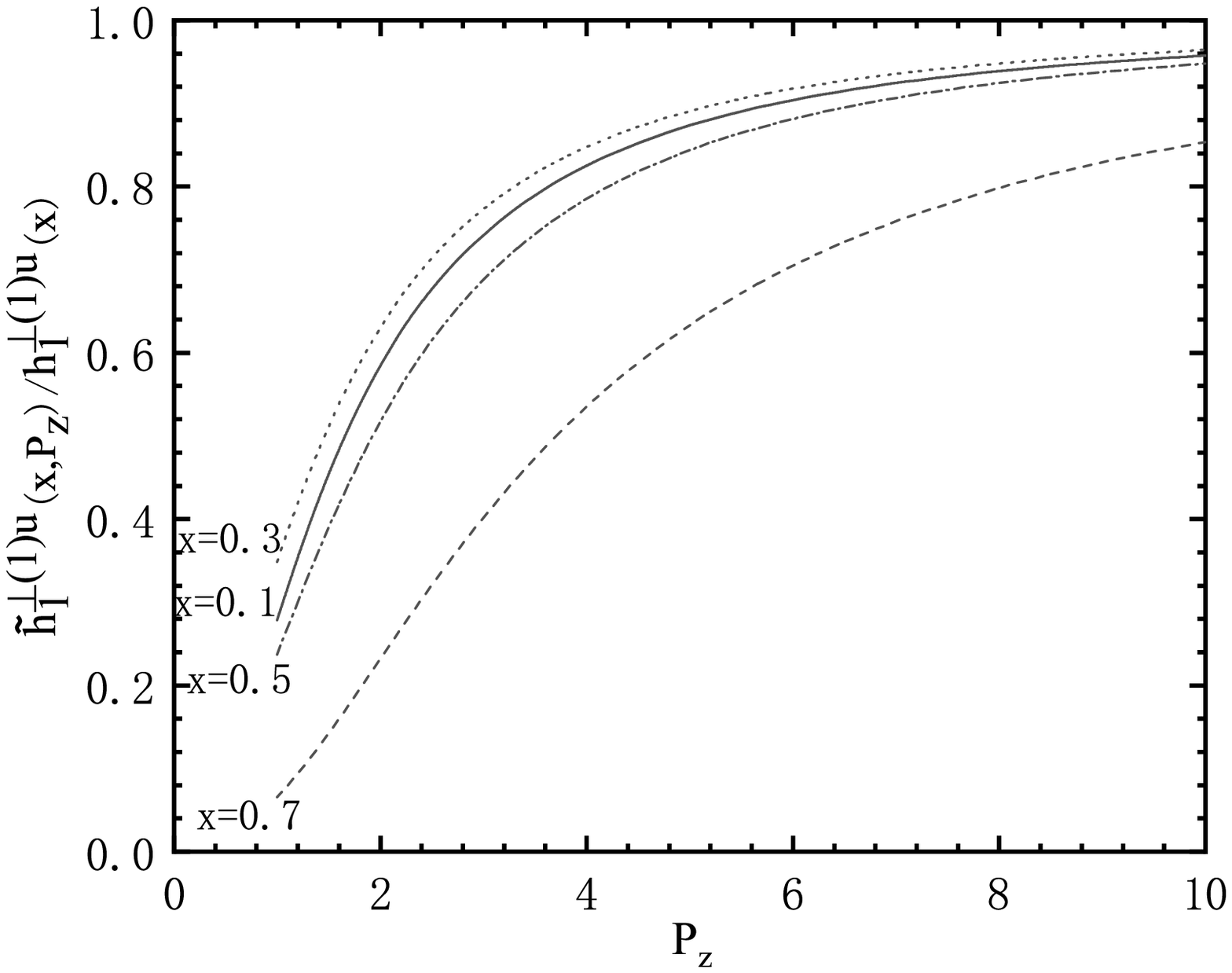}~~~
  \includegraphics[width=0.4\columnwidth]{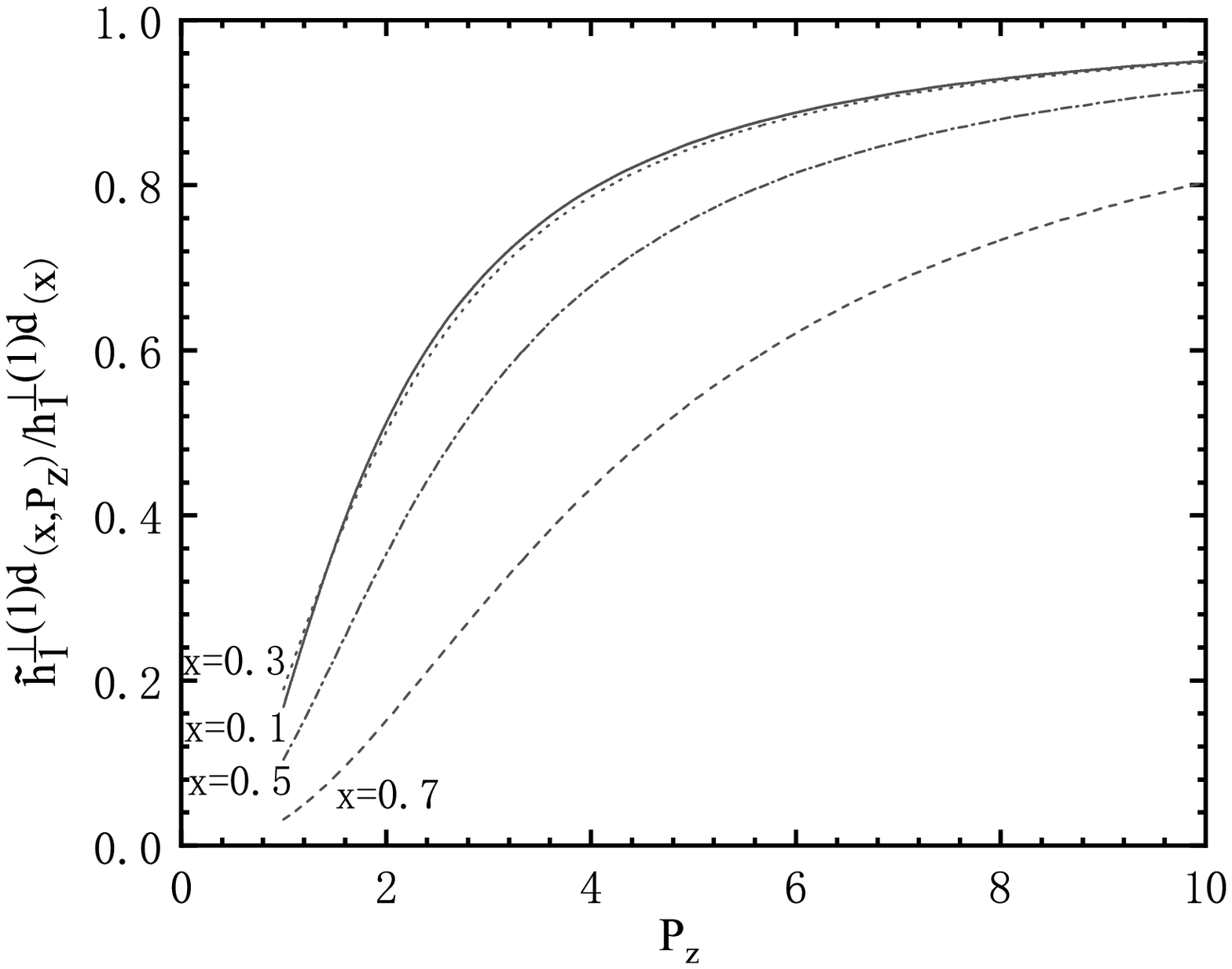}
  \caption{Upper panel: the ratio of $\tilde{f}^{\perp (1)}_{1T}(x,P_z) /{f}^{\perp (1)}_{1T}(x) $ as a function of $P_z$ for the $u$ and $d$ quarks in the spectator model.
  Lower panel: the ratio $\tilde{h}^{\perp (1)}_{1}(x,P_z) /{h}^{\perp (1)}_{1}(x) $ as a function of $P_z$ for the $u$ and $d$ quarks. The solid, dotted, dash-dotted, dashed lines correspond to the results at $x=$0.1, 0.3, 0.5 and 0.7, respectively.}
  \label{fig:zz}
\end{figure}

To provide a more comprehensive discussion on the variation of the quasi-distributions with increasing $P_z$ in different $x$ regions, in Fig.~\ref{fig:zz} we plot the ratios $\tilde{f}^{\perp (1)}_{1T}(x,P_z) /{f}^{\perp (1)}_{1T}(x)$ (upper panel) and $\tilde{h}^{\perp (1)}_1(x,P_z)/{h}^{\perp (1)}_1(x)$ (lower panel) as functions of $P_z$ at fixed $x=$0.1, 0.3, 0.5 and 0.7, respectively.
From the figure one can see that the ratio $\tilde{f}^{\perp (1)}_{1T}(x,P_z) /{f}^{\perp (1)}_{1T}(x)$ is clearly flavor dependent.
For the $u$ quark Sivers distribution the ratio is less than 1 except $x=0.1$; the ratios in different $x$ regions are positive approaches to 1 around $P_z\approx 8$ GeV.
Meanwhile, for the $d$ quark Sivers function, the ratio is negative in the smaller $P_z$  region and turns to be positive in the larger $P_z$ region.
Also, the ratio $\tilde{f}^{\perp (1)}_{1T}(x,P_z) /{f}^{\perp (1)}_{1T}(x)$  for the $d$ quark converges to 1 slower than that for the $u$ quark distribution.
In the case of Boer-Mulders function, the ratios for the $u$ quark and the $d$ quark are very similar, that is, they are positive and less than 1 in the entire $x$ region.
In the region $x<0.5$, the ratio approaches to $0.9-0.95$ at $P_z=8$ GeV, while in the larger $x$ region (such as $x=0.7$) the ratio approaches to $0.8-0.85$ at $P_z=8$ GeV. That is, in the large $x$ region the quasi Boer-Mulders function converges slower.

Compared with the results in Ref.~\cite{Gamberg:2014zwa}, we can find that there are some common features shared by the T-odd quasi-TMDs and the T-even quasi-PDFs.
Firstly, in both cases the quasi-distributions reduce to the standard distributions in the limit $P_z\rightarrow \infty$.
Secondly, the quasi-distributions can have an opposite sign to the standard distributions in certain regions, such as $\tilde{f}_{1T}^{\perp d} (x,k_T^2;P_z)$, $ \tilde{g}_1^d(x,P_z)$ and $\tilde{h}_1^u(x,P_z)$.
However, there is also the feature of the T-odd quasi-TMDs which is different from that of T-even quasi-PDFs.
As is evident from Ref.~\cite{Gamberg:2014zwa}, for the intermediate $x$ region $0.1< x <
0.4 - 0.5$, the quasi-PDFs $\tilde{f}_1^q(x)$, $\tilde{g}_1^q(x)$ and $\tilde{h}_1^q(x)$ approximate the corresponding standard PDFs within $20-30\%$ when $P_z\geq  1.5-2 $ GeV.
While from Fig.~\ref{fig:zz} we find that the T-odd quasi-TMDs in the spectator model are fair approximations to the standard
TMDs (within $20 - 30\%$) in the intermediate $x$ region when $P_z \geq 2.5 - 3$ GeV, which is larger than the case of the T-even quasi-PDFs.
Thus, to obtain the results for the T-odd TMDs as accurate as that for the T-even PDFs in the lattice calculation, one should explore relatively larger $P_z$ region.
Finally, in our study we have chosen the diquark anomalous chromomagnetic moment as $\kappa_a=0$ to simplify the calculation.  
We find that varying $\kappa_a$ between 0 and 1  will not change our numerical result qualitatively, Particularly, in the case $\kappa_a=1$ there is still a fair agreement between the quasi-TMDs and the standard TMDs in the region $P_z \geq 2 - 3$ GeV.

\section{Summary} \label{Sec:5}

In this paper, we computed the two twist-2 T-odd quasi-distributions, the quasi Sivers function $\tilde{f}_{1T}^\perp(x, \bm k_T^2, P_z)$ and the quasi Boer-Mulders function $\tilde{h}_1^\perp(x, \bm k_T^2, P_z)$, in a spectator model
with both scalar diquark and axial-vector diquark.
The quasi-functions are obtained by replacing $\gamma^+$ and $\sigma^{i+}$ with $\gamma_z$ and $\sigma_{iz}$, which make them defined in a four-dimensional Euclidean space rather than in the Minkowski space-time.
We applied the dipolar form factor for the proton-quark-diquark vertex to provide the expressions of the quasi-functions and compare them with the standard functions in the same model.
To perform the integrations over the $l_0$ and $l_z$ components of the gluon four-momentum, we adopted the cut-diagram approach.
We found that the two T-odd quasi-TMDs reduce to the analytical results of the corresponding standard TMDs in the limit $P_z\rightarrow \infty$, which is analogous to the results of the T-even quasi-PDFs $\tilde{f}_1$, $\tilde{g}_1$ and $\tilde{h}_1$.
This observation is also supported by the numerical results for
$\tilde{f}_{1T}^\perp(x, \bm k_T^2; P_z)$ and $\tilde{h}_1^\perp(x, \bm k_T^2; P_z)$ as functions of the transverse momentum $k_T$ at different $x$ and $P_z$.
Another observation is that the convergence depends on $x$, that is, in general the quasi-TMDs approach to the standard ones more quickly in the smaller $x$ region than in the larger $x$ region as $P_z$ increases.
We studied the flavor dependence of the quasi-TMDs and found that
the quasi Sivers functions of the $u$ and $d$ quarks are quite different, while the quasi Boer-Mulders functions are almost flavor blind.
We also calculated the first $k_T$-moment of the T-odd quasi-TMDs $\tilde{f}_{1T}^{\perp(1)}(x, P_z)$ and $\tilde{h}_1^{\perp(1)}(x, P_z)$ as functions of $x$ and $P_z$.
We found that $\tilde{f}_{1T}^{\perp(1)}(x, P_z)$ and $\tilde{h}_1^{\perp(1)}(x, P_z)$  in the spectator model are fair approximations to the standard
ones (within 20-30\%) in the region $0.1<x<0.5$ when $P_z \geq 2.5 -3$ GeV. This is in general larger than the value of the T-even quasi-PDFs $\tilde{f}_1^q(x)$,  $\tilde{g}_1^q(x)$ and $\tilde{h}_1^q(x)$.
In summary, our study has provided model implications and constraints on the quasi Sivers function and the quasi Boer-Mulders function,
and it is possible to access the T-odd standard distributions from lattice QCD calculation in the region $P_z>2.5$ GeV as fair approximations.

\section*{Acknowledgements}
This work is partially supported by the National Natural Science Foundation of China under grant number 12150013.


\begin{thebibliography}{99}

\bibitem{Bacchetta:2006tn}
A.~Bacchetta, M.~Diehl, K.~Goeke, A.~Metz, P.~J.~Mulders and M.~Schlegel,
%``Semi-inclusive deep inelastic scattering at small transverse momentum,''
JHEP \textbf{02}, 093 (2007).

\bibitem{Sivers:1989cc}
  D.~W.~Sivers,
  %``Single Spin Production Asymmetries From The Hard Scattering Of Point-Like
  %Constituents,''
  Phys.\ Rev.\ D {\bf 41}, 83 (1990);
  %%CITATION = PHRVA,D41,83;%%
%\bibitem{Sivers:1990fh}
  % D.~W.~Sivers,
  %``Hard Scattering Scaling Laws For Single Spin Production Asymmetries,''
  Phys.\ Rev.\ D {\bf 43}, 261 (1991).
  %%CITATION = PHRVA,D43,261;%%


\bibitem{Anselmino:2005an}
M.~Anselmino, M.~Boglione, J.~C.~Collins, U.~D'Alesio, A.~V.~Efremov, K.~Goeke, A.~Kotzinian, S.~Menzel, A.~Metz and F.~Murgia, \textit{et al.}
%``Comparing extractions of Sivers functions,''
[arXiv:hep-ph/0511017 [hep-ph]].

\bibitem{Boer:1997nt}
D.~Boer and P.~J.~Mulders,
%``Time reversal odd distribution functions in leptoproduction,''
Phys. Rev. D \textbf{57}, 5780-5786 (1998).

\bibitem{Ji:2013dva}
X.~Ji,
%``Parton Physics on a Euclidean Lattice,''
Phys. Rev. Lett. \textbf{110}, 262002 (2013).

\bibitem{Ji:2014gla}
X.~Ji,
%``Parton Physics from Large-Momentum Effective Field Theory,''
Sci. China Phys. Mech. Astron. \textbf{57}, 1407-1412 (2014).

\bibitem{Lin:2014zya}
H.~W.~Lin, J.~W.~Chen, S.~D.~Cohen and X.~Ji,
%``Flavor Structure of the Nucleon Sea from Lattice QCD,''
Phys. Rev. D \textbf{91}, 054510 (2015).

\bibitem{Alexandrou:2015rja}
C.~Alexandrou, K.~Cichy, V.~Drach, E.~Garcia-Ramos, K.~Hadjiyiannakou, K.~Jansen, F.~Steffens and C.~Wiese,
%``Lattice calculation of parton distributions,''
Phys. Rev. D \textbf{92}, 014502 (2015).

\bibitem{Alexandrou:2016jqi}
C.~Alexandrou, K.~Cichy, M.~Constantinou, K.~Hadjiyiannakou, K.~Jansen, F.~Steffens and C.~Wiese,
%``Updated Lattice Results for Parton Distributions,''
Phys. Rev. D \textbf{96}, no.1, 014513 (2017).

\bibitem{Chen:2016utp}
J.~W.~Chen, S.~D.~Cohen, X.~Ji, H.~W.~Lin and J.~H.~Zhang,
%``Nucleon Helicity and Transversity Parton Distributions from Lattice QCD,''
Nucl. Phys. B \textbf{911}, 246-273 (2016).

\bibitem{Zhang:2017bzy}
J.~H.~Zhang, J.~W.~Chen, X.~Ji, L.~Jin and H.~W.~Lin,
%``Pion Distribution Amplitude from Lattice QCD,''
Phys. Rev. D \textbf{95}, no.9, 094514 (2017).

\bibitem{Zhang:2017zfe}
J.~H.~Zhang \textit{et al.} [LP3],
%``Kaon Distribution Amplitude from Lattice QCD and the Flavor SU(3) Symmetry,''
Nucl. Phys. B \textbf{939}, 429-446 (2019).

\bibitem{Alexandrou:2017huk}
C.~Alexandrou, K.~Cichy, M.~Constantinou, K.~Hadjiyiannakou, K.~Jansen, H.~Panagopoulos and F.~Steffens,
%``A complete non-perturbative renormalization prescription for quasi-PDFs,''
Nucl. Phys. B \textbf{923}, 394-415 (2017).

\bibitem{Chen:2017mzz}
J.~W.~Chen, T.~Ishikawa, L.~Jin, H.~W.~Lin, Y.~B.~Yang, J.~H.~Zhang and Y.~Zhao,
%``Parton distribution function with nonperturbative renormalization from lattice QCD,''
Phys. Rev. D \textbf{97}, no.1, 014505 (2018).

\bibitem{Green:2017xeu}
J.~Green, K.~Jansen and F.~Steffens,
%``Nonperturbative Renormalization of Nonlocal Quark Bilinears for Parton Quasidistribution Functions on the Lattice Using an Auxiliary Field,''
Phys. Rev. Lett. \textbf{121}, no.2, 022004 (2018).

\bibitem{Lin:2017ani}
H.~W.~Lin \textit{et al.} [LP3],
%``Improved parton distribution functions at the physical pion mass,''
Phys. Rev. D \textbf{98}, no.5, 054504 (2018).

\bibitem{Orginos:2017kos}
K.~Orginos, A.~Radyushkin, J.~Karpie and S.~Zafeiropoulos,
%``Lattice QCD exploration of parton pseudo-distribution functions,''
Phys. Rev. D \textbf{96}, no.9, 094503 (2017).

\bibitem{Bali:2017gfr}
  G.~S.~Bali {\it et al.},
  %``Pion distribution amplitude from Euclidean correlation functions,''
  Eur.\ Phys.\ J.\ C {\bf 78}, 217 (2018)
  [arXiv:1709.04325 [hep-lat]].
  %%CITATION = doi:10.1140/epjc/s10052-018-5700-9;%%

\bibitem{Alexandrou:2017dzj}
C.~Alexandrou, S.~Bacchio, K.~Cichy, M.~Constantinou, K.~Hadjiyiannakou, K.~Jansen, G.~Koutsou, A.~Scapellato and F.~Steffens,
%``Computation of parton distributions from the quasi-PDF approach at the physical point,''
EPJ Web Conf. \textbf{175}, 14008 (2018).

\bibitem{Chen:2017gck}
  J.~W.~Chen {\it et al.},
  %``Kaon Distribution Amplitude from Lattice QCD and the Flavor SU(3) Symmetry,''
  arXiv:1712.10025 [hep-ph].
  %%CITATION = ARXIV:1712.10025;%%

\bibitem{Alexandrou:2018pbm}
C.~Alexandrou, K.~Cichy, M.~Constantinou, K.~Jansen, A.~Scapellato and F.~Steffens,
%``Light-Cone Parton Distribution Functions from Lattice QCD,''
Phys. Rev. Lett. \textbf{121}, no.11, 112001 (2018).

\bibitem{Chen:2018xof}
J.~W.~Chen, L.~Jin, H.~W.~Lin, Y.~S.~Liu, Y.~B.~Yang, J.~H.~Zhang and Y.~Zhao,
%``Lattice Calculation of Parton Distribution Function from LaMET at Physical Pion Mass with Large Nucleon Momentum,''
[arXiv:1803.04393 [hep-lat]].

\bibitem{Alexandrou:2018eet}
  C.~Alexandrou, K.~Cichy, M.~Constantinou, K.~Jansen, A.~Scapellato and F.~Steffens,
  %``Transversity parton distribution functions from lattice QCD,''
  arXiv:1807.00232 [hep-lat].
  %%CITATION = ARXIV:1807.11214;%%

\bibitem{Liu:2018uuj}
  Y.~S.~Liu, J.~W.~Chen, L.~Jin, H.~W.~Lin, Y.~B.~Yang, J.~H.~Zhang and Y.~Zhao,
  %``Unpolarized quark distribution from lattice QCD: A systematic analysis of renormalization and matching,''
  arXiv:1807.06566 [hep-lat].
  %%CITATION = ARXIV:1807.06566;%%

\bibitem{Bali:2018spj}
G.~S.~Bali, V.~M.~Braun, B.~Gl\"a\ss{}le, M.~G\"ockeler, M.~Gruber, F.~Hutzler, P.~Korcyl, A.~Sch\"afer, P.~Wein and J.~H.~Zhang,
%``Pion distribution amplitude from Euclidean correlation functions: Exploring universality and higher-twist effects,''
Phys. Rev. D \textbf{98}, no.9, 094507 (2018).

\bibitem{Lin:2018qky}
  H.~W.~Lin, J.~W.~Chen, L.~Jin, Y.~S.~Liu, Y.~B.~Yang, J.~H.~Zhang and Y.~Zhao,
  %``Spin-Dependent Parton Distribution Function with Large Momentum at Physical Pion Mass,''
  arXiv:1807.07431 [hep-lat].
  %%CITATION = ARXIV:1807.07431;%%

\bibitem{LatticeParton:2018gjr}
Y.~S.~Liu \textit{et al.} [Lattice Parton],
%``Unpolarized isovector quark distribution function from lattice QCD: A systematic analysis of renormalization and matching,''
Phys. Rev. D \textbf{101}, no.3, 034020 (2020).

\bibitem{Ji:2018hvs}
X.~Ji, L.~C.~Jin, F.~Yuan, J.~H.~Zhang and Y.~Zhao,
%``Transverse momentum dependent parton quasidistributions,''
Phys. Rev. D \textbf{99}, no.11, 114006 (2019)
[arXiv:1801.05930 [hep-ph]].


\bibitem{Ebert:2019okf}
M.~A.~Ebert, I.~W.~Stewart and Y.~Zhao,
%``Towards Quasi-Transverse Momentum Dependent PDFs Computable on the Lattice,''
JHEP \textbf{09}, 037 (2019)
[arXiv:1901.03685 [hep-ph]].

\bibitem{Ji:2020ect}
X.~Ji, Y.~S.~Liu, Y.~Liu, J.~H.~Zhang and Y.~Zhao,
%``Large-momentum effective theory,''
Rev. Mod. Phys. \textbf{93}, no.3, 035005 (2021)
[arXiv:2004.03543 [hep-ph]].

\bibitem{Ebert:2020gxr}
M.~A.~Ebert, S.~T.~Schindler, I.~W.~Stewart and Y.~Zhao,
%``One-loop Matching for Spin-Dependent Quasi-TMDs,''
JHEP \textbf{09}, 099 (2020)
[arXiv:2004.14831 [hep-ph]].

\bibitem{Ji:2020jeb}
X.~Ji, Y.~Liu, A.~Sch\"afer and F.~Yuan,
%``Single Transverse-Spin Asymmetry and Sivers Function in Large Momentum Effective Theory,''
Phys. Rev. D \textbf{103}, no.7, 074005 (2021)
[arXiv:2011.13397 [hep-ph]].

\bibitem{Gamberg:2014zwa}
L.~Gamberg, Z.~B.~Kang, I.~Vitev and H.~Xing,
%``Quasi-parton distribution functions: a study in the diquark spectator model,''
Phys. Lett. B \textbf{743}, 112-120 (2015)
[arXiv:1412.3401 [hep-ph]].

\bibitem{Bacchetta:2016zjm}
  A.~Bacchetta, M.~Radici, B.~Pasquini and X.~Xiong,
  %``Reconstructing parton densities at large fractional momenta,''
  Phys.\ Rev.\ D {\bf 95}, 014036 (2017)
  [arXiv:1608.07638 [hep-ph]].
  %%CITATION = doi:10.1103/PhysRevD.95.014036;%%

\bibitem{Nam:2017gzm}
  S.~i.~Nam,
  %``Quasi-distribution amplitudes for pion and kaon via the nonlocal chiral-quark model,''
  Mod.\ Phys.\ Lett.\ A {\bf 32}, 1750218 (2017)
  [arXiv:1704.03824 [hep-ph]].
  %%CITATION = doi:10.1142/S0217732317502182;%%

\bibitem{Broniowski:2017wbr}
  W.~Broniowski and E.~Ruiz Arriola,
  %``Nonperturbative partonic quasidistributions of the pion from chiral quark models,''
  Phys.\ Lett.\ B {\bf 773}, 385 (2017)
  [arXiv:1707.09588 [hep-ph]].
  %%CITATION = doi:10.1016/j.physletb.2017.08.055;%%

\bibitem{Hobbs:2017xtq}
  T.~J.~Hobbs,
  %``Quantifying finite-momentum effects in the quark quasidistribution functions of mesons,''
  Phys.\ Rev.\ D {\bf 97}, 054028 (2018)
  [arXiv:1708.05463 [hep-ph]].
  %%CITATION = doi:10.1103/PhysRevD.97.054028;%%

\bibitem{Broniowski:2017gfp}
  W.~Broniowski and E.~Ruiz Arriola,
  %``Partonic quasidistributions of the proton and pion from transverse-momentum distributions,''
  Phys.\ Rev.\ D {\bf 97}, 034031 (2018)
  [arXiv:1711.03377 [hep-ph]].
  %%CITATION = doi:10.1103/PhysRevD.97.034031;%%

\bibitem{Xu:2018eii}
  S.~S.~Xu, L.~Chang, C.~D.~Roberts and H.~S.~Zong,
  %``Pion and kaon valence-quark parton quasidistributions,''
  Phys.\ Rev.\ D {\bf 97}, 094014 (2018)
  [arXiv:1802.09552 [nucl-th]].
  %%CITATION = doi:10.1103/PhysRevD.97.094014;%%

\bibitem{Son:2019ghf}
  H.~D.~Son, A.~Tandogan and M.~V.~Polyakov,
  %``Nucleon quasi-Parton Distributions in the large N$_c$ limit,''
  arXiv:1911.01955 [hep-ph].

\bibitem{Lu:2004au}
Z.~Lu and B.~Q.~Ma,
%``Sivers function in light-cone quark model and azimuthal spin asymmetries in pion electroproduction,''
Nucl. Phys. A \textbf{741}, 200-214 (2004)
[arXiv:hep-ph/0406171 [hep-ph]].

\bibitem{Kang:2010hg}
Z.~B.~Kang, J.~W.~Qiu and H.~Zhang,
%``Quark-gluon correlation functions relevant to single transverse spin asymmetries,''
Phys. Rev. D \textbf{81}, 114030 (2010).

\bibitem{Bacchetta:2003rz}
A.~Bacchetta, A.~Schaefer and J.~J.~Yang,
%``Sivers function in a spectator model with axial vector diquarks,''
Phys. Lett. B \textbf{578}, 109-118 (2004).

\bibitem{Bhattacharya:2018zxi}
S.~Bhattacharya, C.~Cocuzza and A.~Metz,
%``Generalized quasi parton distributions in a diquark spectator model,''
Phys. Lett. B \textbf{788}, 453-463 (2019).

\bibitem{Bhattacharya:2019cme}
S.~Bhattacharya, C.~Cocuzza and A.~Metz,
%``Exploring twist-2 GPDs through quasidistributions in a diquark spectator model,''
Phys. Rev. D \textbf{102}, no.5, 054021 (2020).

\bibitem{Ma:2019agv}
Z.~L.~Ma, J.~Q.~Zhu and Z.~Lu,
%``Quasiparton distribution function and quasigeneralized parton distribution of the pion in a spectator model,''
Phys. Rev. D \textbf{101}, no.11, 114005 (2020).

\bibitem{Brodsky:2002cx}
  S.~J.~Brodsky, D.~S.~Hwang and I.~Schmidt,
  % Final-state interactions and single-spin asymmetries in semi-inclusive
  % deep inelastic scattering,
  Phys.\ Lett.\ B {\bf 530}, 99 (2002).

\bibitem{Boer:2002ju}
D.~Boer, S.~J.~Brodsky and D.~S.~Hwang,
%``Initial state interactions in the unpolarized Drell-Yan process,''
Phys. Rev. D \textbf{67}, 054003 (2003).

\bibitem{Gamberg:2003ey}
L.~P.~Gamberg, G.~R.~Goldstein and K.~A.~Oganessyan,
%``Novel transversity properties in semiinclusive deep inelastic scattering,''
Phys. Rev. D \textbf{67}, 071504 (2003)
[arXiv:hep-ph/0301018 [hep-ph]].

\bibitem{Lu:2004hu}
Z.~Lu and B.~Q.~Ma,
%``Non-zero transversity distribution of the pion in a quark-spectator-antiquark model,''
Phys. Rev. D \textbf{70}, 094044 (2004)
[arXiv:hep-ph/0411043 [hep-ph]].

\bibitem{Gamberg:2007wm}
L.~P.~Gamberg, G.~R.~Goldstein and M.~Schlegel,
%``Transverse Quark Spin Effects and the Flavor Dependence of the Boer-Mulders Function,''
Phys. Rev. D \textbf{77}, 094016 (2008).

\bibitem{Bacchetta:2008af}
A.~Bacchetta, F.~Conti and M.~Radici,
%``Transverse-momentum distributions in a diquark spectator model,''
Phys. Rev. D \textbf{78}, 074010 (2008).

\bibitem{Pasquini:2010af}
  B.~Pasquini and F.~Yuan,
  %``Sivers and Boer-Mulders functions in Light-Cone Quark Models,''
  Phys.\ Rev.\ D {\bf 81}, 114013 (2010)
  [arXiv:1001.5398 [hep-ph]].
\bibitem{Courtoy:2008vi}
A.~Courtoy, F.~Fratini, S.~Scopetta and V.~Vento,
%``A Quark model analysis of the Sivers function,''
Phys. Rev. D \textbf{78}, 034002 (2008)
[arXiv:0801.4347 [hep-ph]].

  \bibitem{Courtoy:2009pc}
  A.~Courtoy, S.~Scopetta and V.~Vento,
  %``Analyzing the Boer-Mulders function within different quark models,''
  Phys.\ Rev.\ D {\bf 80}, 074032 (2009)
  [arXiv:0909.1404 [hep-ph]].
\bibitem{Yuan:2003wk}
  F.~Yuan,
  %``Sivers function in the MIT bag model,''
  Phys.\ Lett.\ B {\bf 575}, 45 (2003)
  [hep-ph/0308157].

\bibitem{Courtoy:2008dn}
A.~Courtoy, S.~Scopetta and V.~Vento,
%``Model calculations of the Sivers function satisfying the Burkardt Sum Rule,''
Phys. Rev. D \textbf{79}, 074001 (2009)
[arXiv:0811.1191 [hep-ph]].

\bibitem{Jakob:1997wg}
R.~Jakob, P.~J.~Mulders and J.~Rodrigues,
%``Modeling quark distribution and fragmentation functions,''
Nucl. Phys. A \textbf{626}, 937-965 (1997).

\bibitem{Ma:2014jla}
Y.~Q.~Ma and J.~W.~Qiu,
%``Extracting Parton Distribution Functions from Lattice QCD Calculations,''
Phys. Rev. D \textbf{98}, no.7, 074021 (2018).


\end{thebibliography}
\end{document}